\documentclass{article} % For LaTeX2e
\usepackage{iclr2027_conference,times}

\usepackage{amsmath,amsfonts,bm}

\def\secref#1{section~\ref{#1}}
\def\eqref#1{equation~\ref{#1}}
\def\1{\bm{1}}

\DeclareMathAlphabet{\mathsfit}{\encodingdefault}{\sfdefault}{m}{sl}
\SetMathAlphabet{\mathsfit}{bold}{\encodingdefault}{\sfdefault}{bx}{n}

\usepackage{hyperref}
\usepackage{url}
\usepackage{booktabs}
\usepackage{amssymb}
\usepackage[table]{xcolor}
\usepackage{graphicx}
\usepackage{subcaption}
\usepackage{array}
\usepackage{multirow}
\usepackage{tabularx}

\title{Where Do Embodied Decisions Come From?
 Rethinking Latent and Explicit Reasoning}

\author{
\makebox[\textwidth][c]{
\renewcommand{\arraystretch}{1.3}
\begin{tabular}{c}
Yuan Lin, Ziyue Zhou, JinLong Zhao, Pei Liu, Haipeng Liu\thanks{Corresponding author.}, Pan Zhou, Kun Zhan \\
Li Auto Inc. \\
\texttt{liuhaipeng@lixiang.com}
\end{tabular}
}
}

\iclrfinalcopy % Uncomment for camera-ready version, but NOT for submission.
\begin{document}

\maketitle

\begin{abstract}
Chain-of-Thought (CoT) reasoning is increasingly incorporated into Vision-Language-Action (VLA) models for embodied agents. Yet, contrary to its benefits in conventional reasoning benchmarks, CoT can make stronger embodied models worse. This capability-dependent effect raises a fundamental question: \emph{what role should explicit reasoning play in embodied policy execution?} We investigate this question by distinguishing explicit reasoning from latent decision computation, i.e., the perception-grounded computation that directly supports action prediction. Under the standard \textit{think-then-act} (TTA) formulation, we first intervene on the generated CoT while keeping the visual input and model parameters fixed. Replacing the model-generated trace with an incorrect one causes a substantial performance collapse (e.g., Navigation F1 drops from 72.14\% to 11.84\%), providing evidence that explicit CoT has strong causal influence when it lies on the action-generation path. We then introduce \textit{decide-then-explain} (DTE), which reverses the generation order so that the action is predicted before the explanation. Together with Visual Conditional Contribution (VCC) and Reasoning Conditional Contribution (RCC), this formulation allows us to characterize how decision behavior changes when explicit reasoning is placed before versus after the action. Across autonomous driving and robotic manipulation, DTE consistently outperforms both TTA and conventional \textit{no-CoT} baselines, while showing increased reliance on perception-grounded computation. Importantly, TTA and DTE differ in both their training and inference factorization; therefore, we do not interpret this comparison as a pure causal identification of latent reasoning. Instead, a complementary TTA-trained, DTE-inference evaluation further examines whether the observed benefit depends on retraining under the new factorization. Overall, our results suggest that explicit CoT is not necessarily best positioned as an intermediate variable for real-time action generation. For capable embodied agents, it can be more effective when used to shape perception-grounded decision computation during training while remaining downstream of the decision at inference time. Code: \url{https://github.com/ocean-luna/openvla-decide-then-explain}.
\end{abstract}

% ============================================================================
% Conceptual contribution
% Introduce reciprocal-response formulation for WAMs.

% Algorithmic contribution
% Hierarchical response generation with bounded reasoning.

% Learning contribution
% Conditional best-response modeling with flow matching.
% ============================================================================
\section{Introduction}

Chain-of-Thought (CoT) can make stronger Vision-Language-Action (VLA) models worse. This counterintuitive phenomenon emerges across both autonomous driving and robotic manipulation, where explicit reasoning provides useful gains for weaker foundation models but becomes neutral or even detrimental as model capability increases. This observation challenges a common assumption inherited from large language models: that inserting an explicit reasoning trace between perception and action should generally improve decision-making. In embodied settings, however, actions must remain tightly grounded in dynamic visual observations and executed under strict temporal and behavioral constraints. An explicit reasoning trace may therefore provide useful guidance, but it may also introduce an additional dependency between perception and action. This raises a more fundamental question than whether CoT improves accuracy: \emph{where should explicit reasoning be placed in an embodied policy?}

Vision-Language-Action models~\cite{shao2025largevlmbasedvisionlanguageactionmodels, Kawaharazuka_2025,li2025surveyvisionlanguageactionmodelsembodied,yu2026surveyefficientvisionlanguageactionmodels} unify visual perception, language understanding, and action generation within a single autoregressive framework, enabling rapid progress in robotic manipulation~\cite{xu2024surveyroboticsfoundationmodels} and autonomous driving~\cite{hu2026visionlanguageactionmodelsautonomousdriving}. Following the reasoning paradigm of large language models, many recent VLA systems adopt a \textit{think-then-act} (TTA) factorization~\cite{wang2025cot4advisionlanguageactionmodelexplicit, Zawalski24-ecot, huang2025thinkactvisionlanguageactionreasoningreinforced}, in which the model first generates an explicit CoT trace and subsequently predicts an action conditioned on that trace. Although this formulation is natural and has shown promise in embodied tasks, it fundamentally changes the information pathway to the action: generated language becomes an intermediate variable between the current visual observation and the final decision. Whether this intermediate variable is beneficial may depend on model capability, reasoning quality, and the degree to which the policy can already perform perception-grounded decision computation without verbalizing it.

This capability-dependent behavior motivates us to distinguish two concepts that are often conflated. We refer to \textit{explicit reasoning} as the verbalized CoT tokens generated by the model, while \textit{latent decision computation} denotes internal computation that supports action prediction without requiring an explicit textual intermediate output. The latter should not be interpreted as evidence that the model possesses a particular form of ``latent reasoning''; rather, it describes the perception-grounded computation available to the policy when explicit language is not placed on the action-generation path. This distinction is important because a comparison between CoT and no-CoT accuracy alone cannot establish where the decision is formed. In particular, \textit{no-CoT} does not constitute a direct measurement of latent decision computation: removing explicit CoT simply removes one observable intermediate representation and may alter the model's computation in multiple ways.

To investigate whether explicit CoT directly influences embodied decisions, we first conduct a targeted intervention under TTA. We replace the model-generated CoT with a semantically incorrect reasoning trace while keeping the visual observation and model parameters fixed. If the generated reasoning were merely an explanatory byproduct, changing its content would not necessarily have a large effect on the subsequent action. Instead, we observe a substantial performance collapse, with Navigation F1 decreasing from 72.14\% to 11.84\%. This intervention provides evidence that explicit CoT has \emph{strong causal influence} on the downstream decision when it is placed on the action-generation path. Importantly, we do not interpret this result as showing that CoT is the sole source of the decision, nor does it establish that the computation occurring without explicit CoT should be identified as ``latent reasoning.'' Rather, it establishes a narrower and more directly testable point: \emph{once explicit CoT is inserted between observation and action, its content can materially alter policy behavior.}

\begin{figure}[t]
    \centering
    \includegraphics[width=0.88\textwidth]{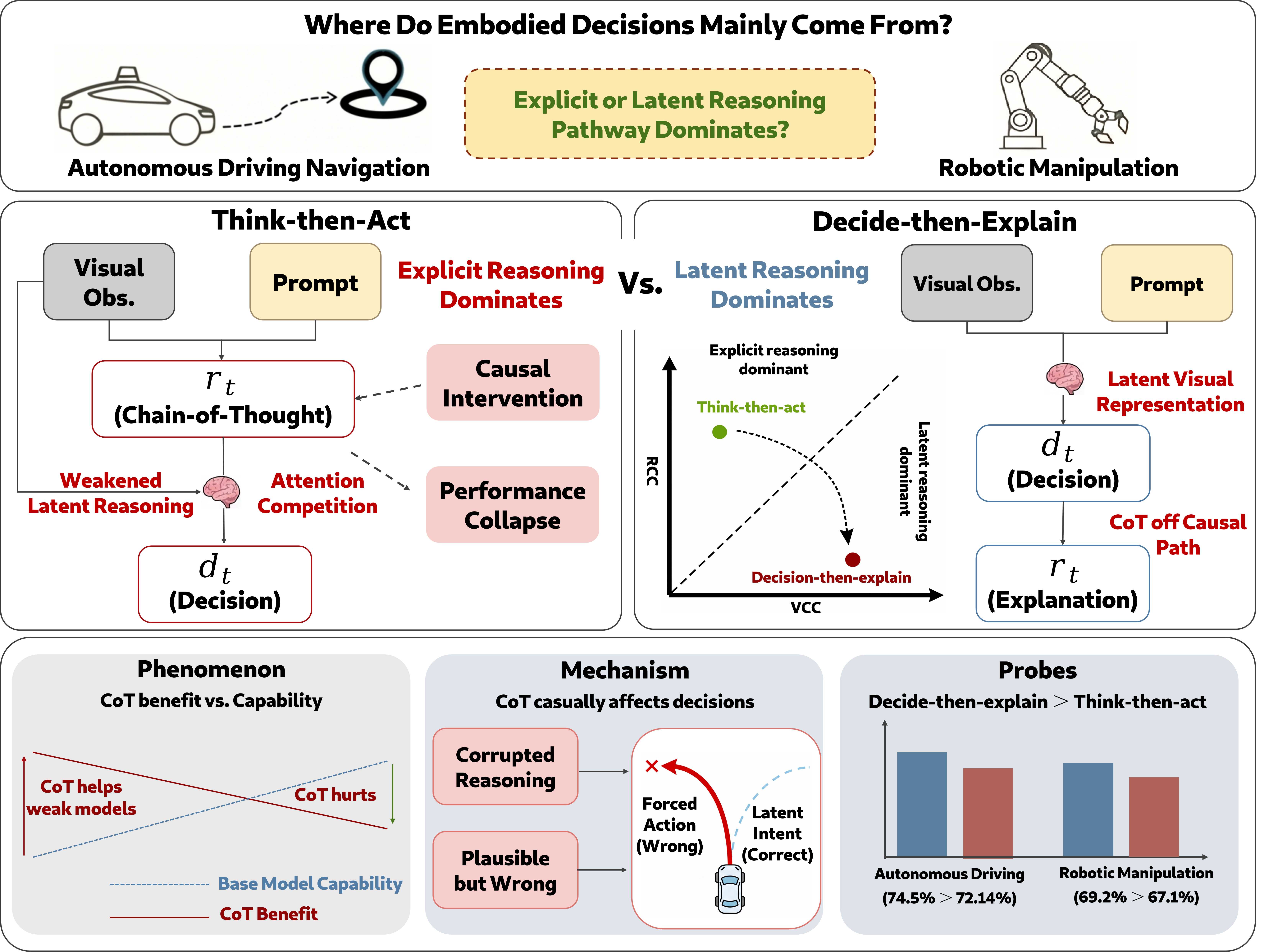}
    \caption{Causal-path view of explicit reasoning in embodied decision-making. Under \textit{think-then-act} (TTA), the generated CoT is placed between visual observations and action generation, allowing its content to exert strong causal influence on the downstream decision. Under \textit{decide-then-explain} (DTE), the action is generated before the explanation, placing explicit CoT downstream of the decision. TTA and DTE differ in both training and inference factorization; therefore, the comparison is treated as a structural probe rather than a pure causal identification of the underlying decision computation.}
    \label{fig:intro}
\end{figure}

The causal influence of CoT also motivates a complementary question: can the policy avoid this dependency while retaining the ability to produce explicit explanations? To study this question, we introduce \textit{decide-then-explain} (DTE), which reverses the conventional generation order. Under DTE, the model first predicts the action and subsequently generates an explanation. This changes the location of explicit reasoning relative to the decision: CoT remains available as an output, but it is no longer an input to the action prediction. Figure~\ref{fig:intro} illustrates the resulting contrast between the two autoregressive factorizations. DTE therefore provides a structural way to study the consequences of placing explicit reasoning downstream of the decision rather than upstream of it.

A crucial qualification is that TTA and DTE differ not only at inference time but also in their training factorization. Consequently, a direct comparison between TTA-trained and DTE-trained models cannot by itself isolate the effect of inference-time causal ordering. We therefore treat DTE as a \emph{structural probe}, rather than as a clean causal intervention on a fixed policy. To further examine whether the observed behavior depends on retraining under DTE, we additionally evaluate a \textit{TTA-trained $\rightarrow$ DTE-inference} setting, in which a model trained with the conventional factorization is evaluated with the decision generated before the explanation. This cross-factorization experiment provides a complementary test of whether moving explicit reasoning downstream can affect policy execution even without retraining the model under the DTE objective.

To characterize the resulting decision pathways, we introduce Visual Conditional Contribution (VCC) and Reasoning Conditional Contribution (RCC). These metrics quantify changes in the action distribution under targeted visual and reasoning interventions, allowing us to measure the relative contribution of perception-grounded computation and explicit reasoning without equating either quantity with ``reasoning'' itself. In particular, RCC is interpreted as a measure of the sensitivity of decisions to explicit reasoning under a given factorization, rather than as a direct measurement of the amount of reasoning performed by the model. When the relevant pathway is structurally absent, the corresponding contribution is treated as \textit{N/A} rather than as zero, since the absence of an intervention pathway does not constitute empirical evidence of zero contribution.

Beyond pathway structure, we further examine the \textit{quality} of explicit CoT. Rather than treating CoT as a binary presence-or-absence variable, we consider CoT quality as an independent axis that can influence its usefulness to the policy. We also introduce controls using random or irrelevant reasoning traces to distinguish the effect of merely inserting additional language from the effect of semantically meaningful reasoning. Together with visual $\times$ CoT interventions, these analyses allow us to test whether the observed policy sensitivity is attributable to the semantic content of the reasoning trace and how that sensitivity interacts with visual evidence.

We find that the role of explicit reasoning is strongly capability- and quality-dependent. For weaker models, CoT can provide useful intermediate guidance. For stronger models, however, placing explicit CoT before the action can become detrimental, and moving the explanation after the decision consistently improves performance across autonomous driving and robotic manipulation. The corresponding increase in VCC indicates greater reliance on perception-grounded computation, while the intervention results show that the effect is not simply attributable to the presence of additional generated tokens. Importantly, these results do not imply that we have identified a unique ``latent reasoning'' mechanism. Instead, they support a more precise conclusion: \emph{for capable embodied policies, explicit reasoning need not be an inference-time causal bottleneck between perception and action.}
Our main contributions and findings are summarized as follows:

\begin{itemize}
    \item \textbf{The CoT Paradox in Embodied Agents.}
    We identify a capability-dependent effect of explicit reasoning across autonomous driving and robotic manipulation: CoT benefits weaker foundation models but can become neutral or harmful for stronger ones. We further show that CoT quality constitutes an important axis of this behavior, indicating that the effect of explicit reasoning cannot be reduced to a simple CoT-versus-no-CoT comparison.

    \item \textbf{Strong Causal Influence of Explicit CoT.}
    Through controlled inference-time interventions under \textit{think-then-act}, we show that changing the semantic content of the generated reasoning can substantially alter downstream actions. Corrupting the reasoning trace causes a catastrophic performance drop, with Navigation F1 decreasing from 72.14\% to 11.84\%, providing evidence that explicit CoT has strong causal influence when placed on the action-generation path.

    \item \textbf{Positioning Explicit Reasoning Downstream of the Decision.}
    We introduce \textit{decide-then-explain} as a structural probe that places explanation after action generation. Although TTA and DTE differ in both training and inference factorization, DTE consistently improves policy performance and increases reliance on perception-grounded computation. A complementary TTA-trained $\rightarrow$ DTE-inference evaluation further tests the role of inference-time ordering without retraining under DTE.

    \item \textbf{Quantifying Decision Pathways.}
    We introduce Visual Conditional Contribution (VCC) and Reasoning Conditional Contribution (RCC) to quantify the sensitivity of action predictions to visual and explicit reasoning information. Combined with random/irrelevant CoT controls and visual $\times$ CoT interventions, these metrics provide a systematic characterization of how explicit reasoning interacts with perception-grounded decision computation.
\end{itemize}

Taken together, our results suggest that the central issue is not whether an embodied model ``has reasoning,'' but rather \emph{where explicit reasoning should be placed within the policy}. For capable VLA agents, explicit CoT can exert strong causal influence when placed before the action, yet making it an obligatory intermediate variable can be counterproductive. Our findings instead support positioning explicit reasoning primarily as a training-time signal that can shape perception-grounded decision computation, while allowing the action to be generated without conditioning on a self-generated reasoning trace at inference time.
    
% \begin{figure*}[htbp]
%     \centering
%     \includegraphics[width=0.88\textwidth]{figures/main.jpg}
%     \caption{Mechanistic study of explicit and latent reasoning in embodied decision-making. We contrast two autoregressive factorizations as structural probes: under \textit{think-then-act}, CoT tokens lie on the causal path to the decision, making corrupted reasoning collapse performance; under \textit{decide-then-explain}, the decision precedes the explanation, removing CoT from the causal path and forcing reliance on latent visual representations.}
% \label{fig:intro}
% \vspace{-1em}
% \end{figure*}
%===============================================================================

\section{Related Works}
\label{sec:citations}

\subsection{Vision-Language-Action Models}

	The VLA paradigm~\cite{shao2025largevlmbasedvisionlanguageactionmodels,Kawaharazuka_2025, li2025surveyvisionlanguageactionmodelsembodied,zhang2025purevisionlanguageaction,poria202510openchallengessteering,yu2026surveyefficientvisionlanguageactionmodels} unifies perception, language understanding, and action generation within a single autoregressive framework. Most VLA systems adopt a Vision-Language Model (VLM)~\cite{lin2025visionlanguagemodelssurvey,zhang2024visionlanguagemodelsvisiontasks,bai2025qwen25vltechnicalreport,liu2023visualinstructiontuning,wang2025internvl35advancingopensourcemultimodal} backbone to directly map visual observations and textual prompts to discrete or continuous control outputs.

	Recent work explores diverse architectural designs for embodied decision-making. In robotics, $\pi_0$~\cite{black2026pi0visionlanguageactionflowmodel} and $\pi_{0.5}$~\cite{intelligence2025pi05visionlanguageactionmodelopenworld} propose generalist VLA policies capable of long-horizon manipulation across heterogeneous domains. VLA-0~\cite{goyal2025vla0} directly predicts actions as text tokens without architectural modification, while LLaDa-VLA~\cite{wen2025lladavlavisionlanguagediffusion} incorporates diffusion-based trajectory modeling into a VLM backbone. In autonomous driving, OpenDriveVLA~\cite{zhou2025opendrivevlaendtoendautonomousdriving} and ORION~\cite{fu2025orionholisticendtoendautonomous} align language-conditioned reasoning with structured motion generation in end-to-end systems. 

	Despite architectural diversity, most VLA models share a common autoregressive formulation in which reasoning tokens are generated before action tokens, adopting a think-then-act sequence. However, the functional role of these reasoning tokens in the underlying decision computation remains underexplored. Our work builds upon unified VLA architectures but shifts the focus from architectural design to the structural role of reasoning within autoregressive decision generation.

\subsection{Explicit and Latent Reasoning in LLMs and MLLMs}

	Recent work on LLMs and MLLMs increasingly distinguishes between \emph{explicit} and \emph{latent} forms of reasoning. Explicit reasoning is verbalized as CoT tokens in the output sequence, and remains the dominant approach across many paradigms because it improves interpretability and can provide useful intermediate supervision~\cite{Zawalski24-ecot,wang2024drivecotintegratingchainofthoughtreasoning,kim24openvla,wang2025cot4advisionlanguageactionmodelexplicit,huang2025thinkactvisionlanguageactionreasoningreinforced,nvidia2026alpamayor1bridgingreasoningaction}. By contrast, latent reasoning refers to computation carried by internal hidden-state dynamics without requiring a fully verbalized textual trace. Recent surveys argue that substantial reasoning may already be internalized within model representations rather than fully expressed in natural language~\cite{li2025implicitreasoninglargelanguage,zhu2025surveylatentreasoning}. This perspective has motivated a growing family of methods that compress, hide, or partially replace verbalized CoT with latent computation, including compressed CoT~\cite{cheng2024compressedchainthoughtefficient}, hidden CoT decoding~\cite{liu2024expeditingelevatinglargelanguage}, latent-token computation~\cite{sun2025enhancinglatentcomputationtransformers,zhang2025latenttokensthinkcausal}, mixed latent-text reasoning~\cite{su2025tokenassortedmixinglatent}, learnable visual tokens for CoT~\cite{qin2025chainofvisualthoughtteachingvlmsthink}, and internal multimodal reasoning in latent space~\cite{liu2025reasoningminddynamicmultimodal}.

	Beyond proposing alternative reasoning approaches, a more recent line of work asks a deeper mechanistic question: whether explicit CoT is the \emph{cause} of the final answer or merely a \emph{verbalization} of computations already formed in hidden states~\cite{dong2025emergentresponseplanningllms,Pal_2023,liu2026largereasoningmodelsnot,wang2026latentchainofthoughtplanningdecoupling}. This distinction becomes especially important in embodied settings, where decisions must remain tightly grounded in visual observations and action dynamics. Several VLA studies have introduced latent planning or latent reasoning components~\cite{huang2025thinkactvisionlanguageactionreasoningreinforced,huang2026fastthinkactefficientvisionlanguageactionreasoning,bai2026latentreasoningvlalatent}. However, they do not directly disentangle whether decisions are causally driven by generated CoT or formed primarily through latent visual representations. Most prior work, therefore, evaluates reasoning mainly through end-task performance, leaving the relative roles of an \emph{explicit reasoning path} and a \emph{latent perception-grounded path} insufficiently resolved. Our work builds on this literature by explicitly distinguishing these two pathways and by using generation order, causal intervention, and contribution metrics to probe which path actually dominates embodied decision-making.

%===============================================================================

\section{Methodology}
\label{sec:Methodology}

	\subsection{Embodied Decision-Making Formulation}
	We consider an embodied decision-making problem in which, at time step $t$, the model receives visual observations $v_t$ and linguistic prompts $p_t$, and produces a decision $d_t$. The decision may be a discrete command (e.g., navigation direction) or a continuous output sequence (e.g., waypoints). The mechanistic distinction of interest is whether this decision is computed primarily through \emph{latent reasoning} internal to the hidden state, or through an \emph{explicit reasoning} variable $r_t$ expressed as CoT tokens:
	\begin{equation}
        \begin{aligned}
        p(d_t \mid v_t, p_t)
        &\quad \text{(latent reasoning only)},
        \text{or}\quad
        p(d_t, r_t \mid v_t, p_t)
        &\quad \text{(with explicit reasoning)}.
        \end{aligned}
    \end{equation}

	Our goal is not to introduce a new decision architecture, but to identify which pathway actually dominates decision-making.

	\subsection{Generation Paradigms as Structural Probes}
	\label{sec:paradigms}
	We treat generation order as a probe of the reasoning mechanism. In the conventional \textit{think-then-act} factorization, the model first generates verbalized reasoning tokens and then produces the decision conditioned on them:
	\begin{equation}
	p(r_t, d_t \mid v_t, p_t)
	= p(r_t \mid v_t, p_t)\, p(d_t \mid r_t, v_t, p_t).
	\end{equation}
	In this factorization, $r_t$ lies on the causal path to $d_t$: it is generated before the decision and directly conditions the computation of $d_t$. Any bias, noise, or semantic error in the reasoning trace can therefore directly perturb the decision distribution. 

	To test whether decisions can instead be made directly from latent visual representations, we use decide-then-explain as a structural probe. Reversing the generation order excises CoT from the causal path of $d_t$ and forces the decision to be formed before any explicit explanation is available:
	\begin{equation}
	p(d_t, r_t \mid v_t, p_t)
	= p(d_t \mid v_t, p_t)\, p(r_t \mid d_t, v_t, p_t).
	\end{equation}
	Here, $d_t$ depends only on visual observations and task prompts; the reasoning sequence is generated afterward as an explanation of an already committed decision. In this sense, think-then-act and decide-then-explain act as switches between explicit-dominant and latent-dominant decision pathways.

	\paragraph{Training-Time Mechanism.}
	Under teacher forcing, the two factorizations induce different decision-learning objectives during training. Think-then-act optimizes the decision under reasoning-conditioned context, $\nabla_\theta \log p(d_t \mid v_t, p_t, r_t)$, whereas decide-then-explain optimizes the decision directly from observations and prompts, $\nabla_\theta \log p(d_t \mid v_t, p_t)$. This distinction does not by itself prove interference inside the network, but it provides a training-time perspective on why CoT can be useful as supervision, while it may be harmful when placed on the inference-time causal path (full derivation in Appendix~\ref{sec:gradient_analysis}).

	\paragraph{Inference-Time Mechanism.}
	At inference, the generated reasoning sequence $r_{1:T_r}$, where $T_r$ denotes the number of reasoning tokens, lies inside the causal receptive field of $d_t$ under think-then-act, so that $r_t$ directly influences $d_t$ via the attention mechanism. Under decide-then-explain, $d_t$ is generated before $r_t$, so reasoning tokens are outside its receptive field.

%===============================================================================
	\subsection{Metrics for Contribution}
	\label{sec:Metrics}

	While paradigm comparison answers whether CoT is structurally in the causal path, it does not quantify how much each modality, visual observations, linguistic prompts, explicit reasoning, actually drives decisions. We introduce contribution metrics ($c_v$, $c_r$) measuring decision probability shifts under targeted modality ablations, complemented by attention ratios that characterize computational focus; all metrics are computed against ground-truth decisions.

	\subsubsection{Visual and Reasoning Conditional Contribution}
	\label{sec:VCC_RCC}

	We define Visual VCC and RCC to quantify modality-specific influence on decision confidence:

	\begin{equation}
	\begin{aligned}
	c_v &= \log p(d_t \mid v_t, p_t) - \log p(d_t \mid \tilde{v}_t, p_t), \\
	c_r &= \log p(d_t \mid v_t, p_t, r_t) - \log p(d_t \mid v_t, p_t, \tilde{r}_t),
	\end{aligned}
	\end{equation}

	where $\tilde{v}_t$ is the ablated visual input, $\tilde{r}_t$ is the ablated reasoning, $(v_t, p_t, d_t) \sim D$, and dataset-level means $\bar{c}_v = \mathbb{E}_D[c_v]$, $\bar{c}_r = \mathbb{E}_D[c_r]$ are reported.

	\paragraph{Implementation.} Visual ablation replaces the image with a black frame, preserving token structure while removing visual content. Reasoning ablation replaces content between \texttt{<think>...</think>} tags with equal-length padding tokens, targeting semantic contribution while preserving sequence structure (details in Appendix~\ref{sec:framework_details}).

	\paragraph{Interpretation.} Positive $c_v$ indicates visual input increases decision confidence; $c_v \approx 0$ suggests weak visual grounding. For $c_r$: positive values indicate reasoning supports the decision, near-zero values indicate reasoning is unused, and negative values indicate conflict.

	\subsubsection{Attention Ratio}
	\label{sec:AttentionRatio}

	To characterize how the model allocates computational focus during decision generation, we analyze attention weights across token groups. Let $A_t^{l,h} \in \mathbb{R}^{|x|}$ denote the attention distribution of head $h$ in layer $l$ when generating token $t$, where $x$ is the full input sequence. We partition $x$ into three disjoint sets: visual tokens $V$, linguistic prompt tokens $P$, and previously generated tokens $G$, with $V \cup P \cup G = x$. The attention ratio for each group is defined by averaging across all layers and heads:
	\begin{equation}
	\begin{aligned}
	r_V = \frac{1}{LH} \sum_{l,h} \sum_{i \in V} A_t^{l,h}(i), \\
	r_P = \frac{1}{LH} \sum_{l,h} \sum_{i \in P} A_t^{l,h}(i), \\
	r_G = \frac{1}{LH} \sum_{l,h} \sum_{i \in G} A_t^{l,h}(i),
	\end{aligned}
	\end{equation}
	where $L$ and $H$ are the number of layers and heads, respectively. By construction (softmax normalization), $r_V + r_P + r_G = 1$.

	\paragraph{Interpretation.}
	These ratios serve as descriptive indicators of attention allocation patterns across paradigms, complementing the contribution metrics by revealing where the model's computational focus shifts when CoT is added or removed. Limitations of attention-based analysis are discussed in Appendix~\ref{sec:framework_details}.

\section{Experiments}
\label{sec:Experiments}

	\subsection{Datasets and Tasks}

	We first study two embodied tasks grounded in real-world autonomous driving data collected from Li Auto user scenarios: \textbf{Navigation Decision} (directional commands) and \textbf{Target Path Generation} (waypoint sequences for approaching a designated target in open-road scenarios). Full task descriptions, dataset variants, reproducibility scope, and statistics are in Appendix~\ref{sec:dataset_setup}.

	\subsection{Experiment Setup}

	Task performance is evaluated by Navigation F1-score (Navigation Decision) and ADE/FDE (Target Path Generation); internal reasoning analysis uses VCC ($c_v$), RCC ($c_r$), and attention ratios ($r_V, r_P, r_G$) as defined in Section~\ref{sec:Metrics}. All models share a SigLIP encoder + Qwen2.5-7B backbone + MLP projector architecture, trained through a staged curriculum (pre-aligned $\to$ AD-general $\to$ object-specific $\to$ task-specific); full architecture and hyperparameter details are in Appendix~\ref{sec:training_detail}. All reported values are test-set averages over held-out splits fully disjoint from training data. Standard deviations for performance metrics are: ±0.10 for F1-score and ±0.01 for ADE/FDE.

	\subsection{Main Results}

		\subsubsection{CoT Effectiveness Inversely Correlates with Foundation Model Capability}

Table~\ref{tab:CoT_comparison} reveals a capability-dependent paradox: as foundation models strengthen, the performance gain from CoT diminishes and eventually turns negative. Since our models share identical architectures and differ only by training stage, this trend isolates latent capability as the driving factor. Weak models (Pre-aligned) benefit significantly from CoT, whereas stronger models (AD-general and Object-specific) degrade.
This reversal implies a compensatory mechanism. When latent representations are weak, explicit reasoning provides necessary scaffolding. However, once latent representations mature, placing CoT on the causal path introduces a linguistic bottleneck that interferes with visually grounded computation (Sec.~\ref{sec:paradigms}). This empirical paradox motivates our central mechanistic inquiry: under \textit{think-then-act}, does CoT causally dictate the decision, or merely verbalize it?

\begin{table}[h]
        \vspace{-1em}
		\centering
		\caption{Comparison of various foundation models with and without CoT.}
		\label{tab:CoT_comparison}
        \scalebox{0.9}{
		\begin{tabular}{llccc}
		\toprule
		\textbf{Foundation} & \textbf{Strategy} & \textbf{Nav. F1 (\%) $\uparrow$} & \textbf{ADE (m) $\downarrow$} & \textbf{FDE (m) $\downarrow$} \\
		\midrule
		\multirow{2}{*}{Object-specific} & w/o CoT & \textbf{73.49} & \textbf{1.82} & \textbf{3.16} \\
		~			& w/ CoT  & 72.14          & 1.83          & 3.17          \\
		\midrule
		\multirow{2}{*}{AD-general}      & w/o CoT & \textbf{70.40} & \textbf{1.75} & \textbf{3.03} \\
						& w/ CoT  & 69.13          & 1.87          & 3.24          \\
		\midrule
		\multirow{2}{*}{Pre-aligned}     & w/o CoT & 63.73          & 2.23          & 3.93          \\
						& w/ CoT  & \textbf{68.89} & \textbf{1.95} & \textbf{3.40} \\
		\bottomrule
		\end{tabular}}
        \vspace{-1em}
\end{table}

		\subsubsection{Causal Role of CoT Content in Decision-Making}

		The capability-dependent pattern above establishes a phenomenon but not yet a mechanism. We further intervene directly on the reasoning trace at inference time while keeping all other inputs and weights fixed. Full details are in Appendix~\ref{sec:causal_interference_details}.
		Two interventions are applied (Table~\ref{tab:causal_interference}). First, replacing the model's CoT with a fixed semantically-mismatched trace collapses Navigation F1 from 72.14\% to 11.84\%, confirming that explicit reasoning is an active causal controller rather than a passive explanation. Second, to rule out simple lexical matching, we use the model's own scene-grounded CoT but randomly invert its directional decision tokens (\emph{plausible-but-wrong} CoT); F1 collapses to 16.8\% with comparable degradation on Target Path Generation. The near-identical collapse across both interventions confirms that the model follows the \emph{reasoning conclusion} in the CoT, not surface lexical patterns.

		\begin{table*}
            
    		\centering
    		\caption{Effect of CoT content corruption on decision performance under think-then-act inference (Object-specific model). Corrupted CoT injects a fixed semantically mismatched trace; Plausible-but-wrong CoT uses the model's own scene-grounded trace with directional tokens randomly inverted.}
    		\label{tab:causal_interference}
    		% \small
    		% \setlength{\tabcolsep}{10pt}
    		% \renewcommand{\arraystretch}{1.1}
    		\begin{tabular}{llccc}
    		\toprule
    		\textbf{Foundation} & \textbf{CoT Condition} & \textbf{Nav. F1 (\%) $\uparrow$} & \textbf{ADE (m) $\downarrow$} & \textbf{FDE (m) $\downarrow$} \\
    		\midrule
    		Object-specific & Normal CoT                   & \textbf{72.14} & \textbf{1.83} & \textbf{3.17} \\
    		Object-specific & Corrupted CoT                & 11.84          & 2.00          & 3.48          \\
    		Object-specific & Plausible-but-wrong CoT      & 16.80          & 1.88          & 3.27          \\
    		\bottomrule
    		\end{tabular}
		\end{table*}

		\subsubsection{Latent Reasoning Dominates When is Removed from the Causal Path}
\label{sec:dominant_reasoning_modes}

To confirm that placing CoT on the causal path degrades performance, we decouple the pathways using the \textit{decide-then-explain} factorization. By generating the decision before the explanation, CoT can no longer condition the action.
Tables~\ref{tab:generation_paradigm_nav} and~\ref{tab:generation_paradigm_target} validate this structural probe. Under \textit{decide-then-explain}, visual contribution ($c_v$) surges, reasoning contribution ($c_r$) drops to zero, and attention shifts back to visual tokens, yielding consistent performance gains over \textit{think-then-act}. If explicit CoT were truly indispensable, removing it from the inference path would cause a performance drop. Instead, severing the $r_t \rightarrow d_t$ causal link forces the model to rely on perception-grounded latent computation, which improves decision quality.
Notably, \textit{decide-then-explain} also outperforms the \textit{no-CoT} baseline (Appendix~\ref{sec:indirect_learning_ablation}). Since CoT exerts no direct computational influence during inference in this setup, this residual advantage stems from training-time effects: CoT supervision refines the latent representations. Thus, explicit CoT serves optimally as an auxiliary training signal rather than an inference-time reasoning chain.

			\begin{table*}[h]
                
    			\centering
    			\caption{Effect of CoT quality on the performance of the object-specific foundation model.}
    			\label{tab:CoT_Quality}
    			\small
    			\setlength{\tabcolsep}{10pt}
    			\renewcommand{\arraystretch}{1.1}
    			\begin{tabular}{llccc}
    			\toprule
    			\textbf{Foundation} & \textbf{CoT Type} & \textbf{Nav. F1 (\%) $\uparrow$} & \textbf{ADE (m) $\downarrow$} & \textbf{FDE (m) $\downarrow$} \\
    			\midrule
    			Object-specific & Without CoT           & 73.49 & 1.82 & 3.16 \\
    			Object-specific & With original CoT     & 72.14 & 1.83 & 3.17 \\
    			Object-specific & With improved CoT     & \textbf{73.85} & \textbf{1.74} & \textbf{2.99} \\
    			\bottomrule
    			\end{tabular}
			\end{table*}

		\subsubsection{CoT Benefit Is Quality-Dependent: Strong Models Require High-Quality Reasoning}

		Explicit reasoning is not uniformly harmful on the decision path. Table~\ref{tab:CoT_Quality} shows that stronger models can still benefit if the CoT is improved. A refined CoT, designed to be concise and causally organized (Appendix~\ref{sec:improved_cot_details}), successfully recovers performance on the object-specific model.
However, this nuance does not invalidate our main finding. While performance degradation under \textit{think-then-act} partly stems from reasoning quality, curating flawless CoT is expensive, difficult to scale, and fails to resolve the structural vulnerability exposed by our causal intervention. Therefore, our core conclusion holds: rather than engineering increasingly refined CoT, the more robust paradigm for embodied agents is to leverage CoT as an auxiliary training signal while deriving decisions directly from latent visual representations.

		\begin{table*}
    		\centering
    		\caption{Impact of generation paradigms on Navigation Decision (Object-specific backbone). Hybrid models are trained on a 50\% decide-then-explain and 50\% think-then-act mixture.}
    		\label{tab:generation_paradigm_nav}
    		\small
    		\begin{tabular}{llcccc}
    		\toprule
    		\textbf{Train Strategy} & \textbf{Inference Prompt} & \textbf{F1 (\%) $\uparrow$} & $\mathbf{\bar{c_v}}$ & $\mathbf{\bar{c_r}}$ & \textbf{Attn R.} ($r_V/r_P/r_G$) (\%) \\
    		\midrule
    		Think-then-act &  Think-then-act & 72.14 & 0.25 & 1.81 & 17.96 / 34.58 / 47.46 \\
    		Decide-then-explain  & Decide-then-explain  & \textbf{74.50} & 1.27 & 0.00 & 30.15 / 47.27 / 22.59 \\
    		% \addlinespace[0.5em]
            \midrule
    		Hybrid          & Think-then-act & 71.71 & 0.24 & 0.31 & 21.50 / 32.96 / 45.53 \\
    		Hybrid          & Decide-then-explain  & 74.04 & 1.85 & 0.00 & 29.11 / 47.17 / 23.71 \\
    		\bottomrule
    		\end{tabular}
		\end{table*}

		\begin{table*}
    		\centering
    		\caption{Impact of generation paradigms on Target Path Generation (Object-specific backbone).}
    		\label{tab:generation_paradigm_target}
    		\small
    		\begin{tabular}{llccccc}
    		\toprule
    		\textbf{Train Strategy} & \textbf{Inference Prompt} & \textbf{ADE $\downarrow$} & \textbf{FDE $\downarrow$} & $\mathbf{\bar{c_v}}$ & $\mathbf{\bar{c_r}}$ & \textbf{Attn R.} ($r_V/r_P/r_G$) (\%) \\
    		\midrule
    		Think-then-act & Think-then-act & 1.83 & 3.17 & 0.48 & 0.09 & 17.21 / 35.94 / 46.85 \\
    		Decide-then-explain  & Decide-then-explain  & \textbf{1.80} & \textbf{3.13} & 0.62 & 0.00 & 25.38 / 44.07 / 30.56 \\
    		% \addlinespace[0.5em]
            \midrule
    		Hybrid          & Think-then-act & 1.87 & 3.23 & 0.47 & 0.00 & 20.13 / 35.28 / 44.59 \\
    		Hybrid          & Decide-then-explain  & 1.83 & 3.15 & 0.62 & 0.00 & 23.59 / 43.82 / 32.60 \\
    		\bottomrule
    		\end{tabular}
		\end{table*}

%===============================================================================

		\subsubsection{Generalization Across Domains, Model Classes, and Training Regimes}
		\label{sec:cross_domain}
		\paragraph{Generalization Across Embodied Domains.}
To validate transferability beyond autonomous driving, we evaluate OpenVLA~\cite{kim24openvla} on the LIBERO-90 robotic manipulation benchmark~\cite{liu2023liberobenchmarkingknowledgetransfer}, utilizing 3,917 CoT-annotated demonstrations from ECoT-Lite~\cite{chen2025trainingstrategiesefficientembodied} (Appendix~\ref{sec:openvla_details}).

As Table~\ref{tab:openvla_libero90} shows, standard \textit{think-then-act} CoT improves the baseline success rate (62.0\% $\rightarrow$ 67.1\%). Crucially, adopting \textit{decide-then-explain} further boosts performance to 69.2\% (std dev $<$ 0.5\%). This confirms that our findings extend to robotic manipulation, reinforcing our core claim: for tightly perception-grounded tasks, placing explicit reasoning on the inference-time causal path is suboptimal.

			\begin{table*}[h]
                \vspace{-1em}
    			\centering
    			\caption{Generalization to robotic manipulation tasks using OpenVLA on LIBERO-90.}
    			\label{tab:openvla_libero90}
    			\small
    			\begin{tabular}{lcc}
    			\toprule
    			\textbf{Model} & \textbf{Train Strategy} & \textbf{Success Rate (\%) $\uparrow$} \\
    			\midrule
    			OpenVLA Baseline~\cite{belkhale2024minivla} & w/o CoT & 62.0 \\
    			OpenVLA + CoT & Think-then-act & 67.1 \\
    			OpenVLA + CoT & Decide-then-explain & \textbf{69.2} \\
    			\bottomrule
    			\end{tabular}
                
			\end{table*}

		\paragraph{Generalization Across Model Classes and Training Regimes.}
To assess generation-order effects beyond embodied tasks, we evaluate two general-purpose VLMs, Qwen2.5-VL-7B~\cite{bai2025qwen25vltechnicalreport} and Video-R1~\cite{videor1}, in a zero-shot setting. As a supplementary observation, we modify only the output token ordering (keeping parameters fixed) and evaluate on three video understanding benchmarks (MMVU, VSI-Bench, VideoMME; Appendix~\ref{sec:vlm_details}).
Table~\ref{tab:video_r1_zero_shot} shows that \textit{decide-then-explain} prompting consistently improves zero-shot performance. While we restrict our primary mechanistic claims to embodied decision-making, these results suggest that removing CoT from the causal path may offer broader benefits across diverse model classes and training regimes.

		\begin{table*}[h]
			\centering
			\caption{Zero-shot evaluation of generation paradigms on general-purpose video language models.}
			\label{tab:video_r1_zero_shot}
			\small
			\begin{tabular}{llccc}
			\toprule
			\textbf{Model} & \textbf{Benchmark} & \textbf{Think-then-act (\%)} & \textbf{Decide-then-explain (\%)}  \\
			\midrule
			\multirow{3}{*}{Qwen2.5-VL-7B}  & MMVU        & 61.59          & \textbf{63.84}  \\
						& VSI-Bench   & 32.41          & \textbf{35.26} \\
						& VideoMME    & 50.93          & \textbf{54.51} \\
			% \addlinespace[0.5em]
            \midrule
			\multirow{3}{*}{Video-R1}       & MMVU        & 59.20          & \textbf{63.04}  \\
						& VSI-Bench   & 29.75          & \textbf{30.84}  \\
						& VideoMME    & 51.00          & \textbf{55.07} \\
			\bottomrule
			\end{tabular}
		\end{table*}

\section{Conclusion and Limitations}
\label{sec:conclusion}

We mechanistically investigated explicit and latent reasoning in embodied decision-making. Addressing the paradox that CoT degrades stronger models, we revealed a shift in pathway dominance. Under \textit{think-then-act}, CoT causally dictates decisions; injecting corrupted reasoning collapses performance. Conversely, structurally removing CoT from the inference path (\textit{decide-then-explain}) eliminates reasoning reliance ($c_r \approx 0$), boosts visual contribution, and improves overall performance. The residual gains over \textit{no-CoT} baselines confirm that CoT provides vital supervision via training-time gradients. Ultimately, capable embodied agents derive decisions primarily from latent visual representations, rendering explicit CoT optimal as an auxiliary training signal rather than an inference-time causal mediator.

Our limitations include: (1) a focus on short-horizon tasks, leaving long-horizon planning dynamics unexplored; (2) defining latent reasoning strictly as the absence of CoT tokens, omitting alternative non-semantic or structured representations; and (3) reliance on proprietary driving data, which may limit the quantitative reproducibility of specific metrics.

% ============================================================================
% ICLR statements
% ============================================================================
\subsection*{AI use statement}

Generative AI tools were used to assist with language editing, \LaTeX{}
formatting, and code inspection. All technical content, experimental designs,
implementations, results, and claims were reviewed and verified by the authors.
The authors take full responsibility for the final content of this work.

\subsection*{Reproducibility statement}

The model formulation and training objective are described in
\secref{sec:method}. Experimental settings, evaluation metrics, and ablations
are reported in \secref{sec:experiments}; additional implementation details
will be provided in the appendix and supplementary material.

\bibliography{iclr2027_conference}
\bibliographystyle{iclr2027_conference}

\appendix
% ============================================================================
% Appendix
% ============================================================================
\section{Analysis Framework and Optimization Details}
\label{sec:framework_details}
\subsection{Analysis Framework: Implementation Details}

\subsubsection{VCC Implementation Details}

To implement visual ablation without altering token sequence length or positional encodings, we replace the original image with a \textit{content-neutralized placeholder}---a uniformly black frame that removes all scene-specific visual content while preserving the structural properties (token count, positional encodings) required by the autoregressive model. $c_v$ should be interpreted as measuring the \textit{association} between visual presence and decision confidence rather than a strictly causal contribution. We adopt this approach rather than attention masking for two reasons. First, masking visual token attention weights removes visual tokens from the computation graph entirely, conflating the absence of visual \emph{content} with the absence of visual \emph{structure}; content-neutralized replacement removes scene-specific information while preserving structure. Second, this design is symmetric with the RCC ablation, where attention masking is not feasible (see below): both metrics follow the same philosophy of \emph{content-neutral replacement under fixed sequence structure}, ensuring $\bar{c}_v$ and $\bar{c}_r$ measure analogous quantities and remain directly comparable. This approach is analogous to replacing input tokens with \texttt{[MASK]} in standard NLP input perturbation studies: the placeholder is structurally valid but semantically empty, isolating the contribution of visual content. Since $c_v$ is used for \emph{relative comparison across experimental conditions} rather than as an absolute measure, any systematic bias introduced by the out-of-distribution placeholder is constant across conditions and does not affect relative rankings. In practice, this yields consistent relative rankings across all reported experimental conditions (Section~\ref{sec:Experiments}), supporting the validity of cross-condition comparisons.

\subsubsection{RCC Implementation Details}

To ablate reasoning while preserving sequence structure, we replace the content between \texttt{<think>} and \texttt{</think>} tags with padding tokens of equal length. This design targets a specific question: \textit{given that a reasoning sequence has been generated, does the model utilize its semantic content?} Attention masking of reasoning tokens is not feasible in the autoregressive setting: the boundary of the \texttt{<think>...</think>} block is only known after the closing tag has been generated, making it impossible to apply masks during inference. Padding-based ablation is therefore the only practical post-hoc approach: after the full sequence is generated, we replace reasoning content and recompute the decision probability in a single forward pass. Complete token removal would collapse the input to a no-CoT configuration, conflating structural and semantic contributions. Padding-based ablation isolates the semantic contribution of reasoning content under a fixed positional and structural context.

\subsubsection{Attention Ratio: Limitations}

Averaging attention weights across all layers treats each layer equally, though transformer layers encode different levels of abstraction. Furthermore, attention weights reflect where the model looks, not necessarily what information it uses---high attention weight does not always correspond to high causal information flow. Despite these limitations, attention ratios provide a complementary view alongside contribution metrics, particularly for detecting coarse shifts in computational focus across generation paradigms.

\subsubsection{Optimization Perspective: Reasoning-Conditioned vs. Decoupled Decision Learning}
\label{sec:gradient_analysis}

This section provides a training-time optimization perspective complementary to the causal analyses in the main text. We analyze how the choice of autoregressive factorization changes the form of the decision-learning objective under teacher forcing, where ground-truth reasoning tokens are used as input during training. The purpose of this derivation is modest: it clarifies how think-then-act and decide-then-explain couple decision learning to reasoning in different ways. It should be read as a qualitative interpretation of optimization structure, not as a formal proof of internal interference or representation degradation.

\paragraph{Preliminaries: Autoregressive Likelihood.}

Consider a training dataset $\mathcal{D} = \{(v_i, p_i, d_i, r_i)\}_{i=1}^N$ where each sample contains visual observation $v_i$, prompt $p_i$, ground-truth decision $d_i$, and ground-truth reasoning $r_i$. For models with explicit reasoning, the maximum likelihood objective is:
\begin{equation}
\mathcal{L}(\theta) = -\frac{1}{N}\sum_{i=1}^N \log p_\theta(d_i, r_i \mid v_i, p_i).
\end{equation}
The training gradient is $\nabla_\theta \mathcal{L}(\theta) = -\frac{1}{N}\sum_{i=1}^N \nabla_\theta \log p_\theta(d_i, r_i \mid v_i, p_i)$. The critical question is how this gradient decomposes under different autoregressive factorizations.

\paragraph{Case 1: Think-then-Act (Conditioned Decision Gradient).}

Under think-then-act, the joint likelihood factorizes as:
\begin{equation}
p_\theta(d_i, r_i \mid v_i, p_i) = p_\theta(r_i \mid v_i, p_i) \cdot p_\theta(d_i \mid r_i, v_i, p_i).
\end{equation}
The log-likelihood gradient decomposes into two terms:
\begin{equation}
\nabla_\theta \log p_\theta(d_i, r_i \mid v_i, p_i) = \underbrace{\nabla_\theta \log p_\theta(r_i \mid v_i, p_i)}_{\text{reasoning gradient}} + \underbrace{\nabla_\theta \log p_\theta(d_i \mid r_i, v_i, p_i)}_{\text{decision gradient}}.
\label{eq:tta_gradient}
\end{equation}

The decision gradient term is evaluated under the ground-truth reasoning $r_i$. Let $h_\theta(\cdot)$ denote the hidden state after processing the input. The decision probability can be written as:
\begin{equation}
p_\theta(d_i \mid r_i, v_i, p_i) = \text{softmax}\big(W_o h_\theta(r_i, v_i, p_i) + b_o\big)[d_i],
\end{equation}
where $h_\theta(r_i, v_i, p_i)$ is the final hidden state conditioning on all tokens including the reasoning sequence $r_i$. The decision gradient is:
\begin{equation}
\nabla_\theta \log p_\theta(d_i \mid r_i, v_i, p_i) = \frac{\partial \log p_\theta(d_i \mid r_i, v_i, p_i)}{\partial h_\theta} \cdot \frac{\partial h_\theta}{\partial \theta}(r_i, v_i, p_i).
\label{eq:tta_chain_rule}
\end{equation}

Crucially, under think-then-act, the decision objective is optimized \emph{conditional on} the reasoning tokens $r_i$. Thus, the learned decision mapping is not simply from $(v_i, p_i)$ to $d_i$, but from $(r_i, v_i, p_i)$ to $d_i$. This means that decision learning is structurally coupled to the reasoning context during training. On its own, this does not prove that the model necessarily learns a worse visual representation, but it does show that the optimization target differs from direct perception-to-decision learning.

\paragraph{Case 2: Decide-then-Explain (Decoupled Decision Gradient).}

Under decide-then-explain, the joint likelihood factorizes as:
\begin{equation}
p_\theta(d_i, r_i \mid v_i, p_i) = p_\theta(d_i \mid v_i, p_i) \cdot p_\theta(r_i \mid d_i, v_i, p_i).
\end{equation}
The log-likelihood gradient decomposes as:
\begin{equation}
\nabla_\theta \log p_\theta(d_i, r_i \mid v_i, p_i) = \underbrace{\nabla_\theta \log p_\theta(d_i \mid v_i, p_i)}_{\text{decision gradient}} + \underbrace{\nabla_\theta \log p_\theta(r_i \mid d_i, v_i, p_i)}_{\text{reasoning gradient}}.
\label{eq:dte_gradient}
\end{equation}

The decision gradient does not depend on the reasoning sequence:
\begin{equation}
\nabla_\theta \log p_\theta(d_i \mid v_i, p_i) = \frac{\partial \log p_\theta(d_i \mid v_i, p_i)}{\partial h_\theta} \cdot \frac{\partial h_\theta}{\partial \theta}(v_i, p_i).
\label{eq:dte_chain_rule}
\end{equation}
Here $h_\theta(v_i, p_i)$ depends only on visual observations and prompts, not on reasoning tokens. This structural decoupling means that the decision objective is evaluated directly from multimodal inputs without conditioning on a reasoning trace.

\paragraph{Optimization Objective Comparison.}

The fundamental difference manifests in the gradient context through which the decision tokens are optimized. Both paradigms supervise all generated tokens jointly under teacher forcing; the distinction lies solely in what context the decision gradient flows through. Under think-then-act (TTA), the decision token loss is evaluated with a hidden state that includes the reasoning context $h_\theta(r_i, v_i, p_i)$:
\begin{equation}
\mathcal{L}_{\text{TTA}}^{\text{decision}}(\theta) = -\frac{1}{N}\sum_{i=1}^N \log p_\theta(d_i \mid r_i, v_i, p_i).
\end{equation}
The model therefore learns to predict decisions conditioned on the preceding reasoning trace, rather than directly from visual observations alone.

Under decide-then-explain (DTE), the decision token loss is evaluated with a hidden state $h_\theta(v_i, p_i)$ that contains no reasoning context:
\begin{equation}
\mathcal{L}_{\text{DTE}}^{\text{decision}}(\theta) = -\frac{1}{N}\sum_{i=1}^N \log p_\theta(d_i \mid v_i, p_i).
\end{equation}
The model therefore learns to predict decisions directly from visual observations and task prompts, without intermediate conditioning on reasoning.

\paragraph{Connection to Experimental Observations.}

This optimization perspective is consistent with the experimental observation that CoT helps weaker models yet can hurt stronger ones under think-then-act:

\begin{itemize}
    \item \textbf{Weak models (Pre-Aligned):} Their latent representations $h_\theta(v_i, p_i)$ may be insufficient to encode task-relevant information. In this regime, reasoning supervision can provide useful auxiliary structure during training, so reasoning-conditioned learning can still yield a net benefit.

    \item \textbf{Strong models (AD-General, Object-Specific):} Their latent representations already encode richer task-relevant information. In this regime, conditioning decision learning on reasoning tokens may provide diminishing benefit while shifting optimization away from the direct perception-to-decision objective, making interference more likely.

    \item \textbf{Decide-then-explain with strong models:} The decision objective $\log p_\theta(d_i \mid v_i, p_i)$ is structurally decoupled from reasoning. CoT supervision can still help through the auxiliary reasoning term $\nabla_\theta \log p_\theta(r_i \mid d_i, v_i, p_i)$, while inference-time decisions remain outside the direct causal influence of reasoning tokens. This is consistent with the empirical pattern that CoT can retain value as an auxiliary training signal without serving as a direct causal mediator at inference time.
\end{itemize}

Overall, this optimization view supports a limited but useful conclusion: factorization order changes what the model is asked to optimize during decision learning. Combined with the causal results in the main text, this helps explain why CoT may be valuable as an auxiliary training signal yet undesirable as a direct causal mediator at inference time. We do not claim that this derivation alone proves the internal source of performance degradation; rather, it provides a training-time interpretation consistent with the observed evidence.

\section{Dataset Construction and Evaluation Statistics}
\label{sec:dataset_setup}
\subsection{Dataset Setup}

Our dataset is a supervised vision-language-action dataset comprising two embodied decision-making tasks: Navigation Decision and Target Path Generation. Each sample consists of visual observations, task instructions, and action outputs, with CoT annotations. The dataset includes several variants, as detailed below, summarized in Table~\ref{tab:Dataset_Statistics}.

\subsubsection{Task I: Navigation Decision}
The Navigation Decision task focuses on high-level discrete decision-making in unstructured underground environments, comprising 175k VQA samples. The model must map visual observations (e.g., floor signage, zone markers, exit markers) to navigational decisions. We categorize the dataset into four critical sub-tasks in application: (1) Parking exit navigation,
(2) Elevator hall navigation, (3) Area search, (4) Floor search.
\begin{quote}
\textbf{Example (Task I):} \\
\textbf{Instruction:} You are an intelligent driving system. The task is to \textit{exit parking}. Analyze the visual cues and provide the reasoning and decision. Output format: Place the reasoning process in \texttt{<think></think>} and the navigation decision in \texttt{<answer></answer>}. Example: \texttt{<think>}Reasoning\texttt{</think>} \texttt{<answer>}Navigation decision\texttt{</answer>} \\
\textbf{Output:} \\
\texttt{<think>} \\
\textbf{Scene Description:} Key observations relevant to the exit parking task in the current scene include a right-turn exit sign, a ground arrow indicating straight and left-turn in the opposite direction, and an intersection allowing straight and right turns. The area is Zone C1. Several parked vehicles are ahead, with no pedestrians or other obstacles present. \\
\textbf{Analysis:} The scene contains an intersection that permits straight and right turns, while left turns are prohibited. The right-turn exit sign is associated with this intersection, indicating a right turn at this location. \\
\textbf{Decision:}  Based on the right-turn exit sign and intersection analysis, the navigation decision is to turn right. \\
\texttt{</think>} \\
\texttt{<answer>}At \texttt{<12.25, -0.90>}, turn right.\texttt{</answer>}
\end{quote}

\subsubsection{Task II: Target Path Generation}
The Target Path Generation task consists of 160k VQA samples and emphasizes spatial grounding and waypoint prediction in open-road driving scenarios. The model is required to output:
(1) a reasoning process,
(2) an immediate navigation intent,
(3) five future waypoints,
(4) the target 2D bounding box, and
(5) the target's goal coordinates.
During training, future waypoints are discretized using an action tokenizer and represented as special tokens.

\begin{quote}
\textbf{Example (Task II):} \\
\textbf{Instruction:} You are an intelligent driving system capable of analyzing and understanding complex driving scenarios. The task is \texttt{Park near the silver SUV ahead}. Based on the front-view camera image of the ego vehicle, analyze the navigation-relevant environmental information and provide the reasoning process and navigation decision. Output format: place the reasoning process in \texttt{<think></think>}, the navigation decision in \texttt{<answer></answer>}, the five corresponding trajectory points in \texttt{<traj></traj>}, the target's 2D bounding box in \texttt{<box></box>}, and the target's actual world coordinates in \texttt{<goal></goal>}. Example: \texttt{<think>}Reasoning\texttt{</think>} \texttt{<answer>}Navigation decision\texttt{</answer>} \texttt{<traj>}Waypoints\texttt{</traj>} \texttt{<box>}Navigation decision\texttt{</box>} \texttt{<goal>}Target position\texttt{</goal>} \\
\textbf{Output:} \\
\texttt{<think>} \\
\textbf{Scene Description:} The road is flanked by trees with clear lane markings. A black car is on the left, and a silver SUV is parked on the right. \\
\textbf{Decision:} To park near the target SUV, the vehicle will maintain its lane, decelerate, and steer toward the right-front. \\
\texttt{</think>} \\
\texttt{<answer>}Proceed to target.\texttt{</answer>} \\
\texttt{<traj>}\{[[3.44, 0.04], [5.51, -0.35], [7.58, -0.73], [9.66, -1.11], [11.73, -1.5]], Proceed to target\}\texttt{</traj>} \\
\texttt{<box>}[[0.6, 0.71], [0.71, 0.94]]\texttt{</box>} \\
\texttt{<goal>}[12.34, -2.98]\texttt{</goal>}
\end{quote}

\subsubsection{Without CoT Dataset}
In the without-CoT setting, all explicit reasoning annotations are removed, while the input instructions and action supervision remain unchanged.

\begin{quote}
\textbf{Example (Task I without CoT):} \\
\textbf{Instruction:} You are an intelligent driving system. The task is to \textit{exit parking}. Analyze the visual cues and provide the reasoning and decision. Output format: Place the reasoning process in \texttt{<think></think>} and the navigation decision in \texttt{<answer></answer>}. Example: \texttt{<think>}Reasoning\texttt{</think>} \texttt{<answer>}Navigation decision\texttt{</answer>}\\
\textbf{Output:} \\
\texttt{<think></think>} \\
\texttt{<answer>}At \texttt{<12.25, -0.90>}, turn right.\texttt{</answer>}
\end{quote}

\subsubsection{Improved CoT Dataset}
\label{sec:improved_cot}
The improved CoT focuses on concise perception-to-decision causal reasoning. The key design principles are: (1) retain only task-relevant observations that directly influence the decision; (2) reduce scene descriptions and task-irrelevant environmental features; (3) establish a stronger causal mapping between visual cues and actions; (4) remove redundant safety prompts and advice. Concretely, we first extract task-relevant key scenes from each input, then construct concise expert annotations that encode the causal chain from scene perception to decision, and finally use these annotations to guide natural-language CoT generation. This structural design provides high-purity reasoning signals with minimal textual noise. Full generation details and quantitative comparison against Original CoT are provided in Appendix~\ref{sec:improved_cot_details}.

\begin{quote}
\textbf{Example (Task I Improved CoT):} \\
\textbf{Instruction:} You are an intelligent driving system. The task is to \textit{exit parking}. Analyze the visual cues and provide the reasoning and decision. Output format: Place the reasoning process in \texttt{<think></think>} and the navigation decision in \texttt{<answer></answer>}. Example: \texttt{<think>}Reasoning\texttt{</think>} \texttt{<answer>}Navigation decision\texttt{</answer>}\\
\textbf{Output:} \\
\texttt{<think>} \\
\textbf{Scene description:} Key observations relevant to the exit parking task in the current scene include a right-turn exit sign, a ground arrow indicating straight and left-turn in the opposite direction, and an intersection allowing straight and right turns. Zone C1. \\
\textbf{Traffic direction analysis:} The scene contains an intersection where straight and right turns are permitted, while left turns are prohibited. The right-turn exit sign is associated with this intersection.\\
\texttt{</think>} \\
\texttt{<answer>}At \texttt{<12.25, -0.90>}, turn right.\texttt{</answer>}
\end{quote}

\subsubsection{Decide-then-explain Dataset}
In the decide-then-explain setting, the output order is modified such that the decision precedes the reasoning process, while all other content remains identical. This ensures that the explicit reasoning does not condition the decision token generation in an autoregressive manner.

\begin{quote}
\textbf{Example (Task I Decide-then-explain):} \\
\textbf{Instruction:} You are an intelligent driving system. The task is to \textit{exit parking}. Analyze the visual cues and provide the reasoning and decision. Output format: Place the reasoning process in \texttt{<think></think>} and the navigation decision in \texttt{<answer></answer>}. Example: \texttt{<answer>}Navigation decision\texttt{</answer>} \texttt{<think>}Reasoning\texttt{</think>}\\
\textbf{Output:} \\
\texttt{<answer>}At \texttt{<12.25, -0.90>}, turn right.\texttt{</answer>}\\
\texttt{<think>} \\
\textbf{Scene Description:} Key observations relevant to the exit parking garage task in the current scene include a right-turn exit sign, a ground arrow indicating straight and left-turn in the opposite direction, and an intersection allowing straight and right turns. The area is Zone C1. Several parked vehicles are ahead, with no pedestrians or other obstacles present. \\
\textbf{Analysis:} The scene contains an intersection that permits straight and right turns, while left turns are prohibited. The right-turn exit sign is associated with this intersection, indicating a right turn at this location. \\
\textbf{Decision:}  Based on the right-turn exit sign and intersection analysis, the navigation decision is to turn right. \\
\texttt{</think>} \\
\end{quote}

\begin{table}[h]
\centering
\caption{Dataset Statistics and Output Modalities.}
\label{tab:Dataset_Statistics}
\small 
\renewcommand{\arraystretch}{1.3} 
\begin{tabularx}{\textwidth}{l l c >{\raggedright\arraybackslash}X}
\toprule
\textbf{Task} & \textbf{Variant} & \textbf{Samples} & \textbf{Output Modality} \\
\midrule
Navigation & Standard (think-then-act) & 175k & CoT + Decision \\
Decision & Without CoT & 175k & Decision only \\
& Improved CoT & 175k & Improved CoT + Decision \\
& Decide-then-explain & 175k & Decision $\rightarrow$ CoT \\
\midrule
Target Path & Standard (think-then-act) & 160k & CoT + Decision + Path + Bbox + Goal \\
Generation & Without CoT & 160k & Decision + Path + Bbox + Goal \\
& Improved CoT & 160k & Improved CoT + Decision + Path + Bbox + Goal \\
& Decide-then-explain & 160k & Decision $\rightarrow$ CoT + Path + Bbox + Goal \\
\bottomrule
\end{tabularx}
\end{table}

\subsubsection{Evaluation Set Statistics}
\label{sec:stat_sig}

All reported metrics are computed over dedicated held-out test sets that are fully disjoint from the training data.
The Navigation Decision test set comprises approximately 5{,}000 samples spanning the four sub-tasks (parking exit navigation, elevator navigation, area search, and floor search).
The Target Path Generation test set comprises approximately 16{,}000 samples covering open-road and dynamic scenarios.

The small absolute magnitudes of improvement are consistent with the near-ceiling difficulty of well-trained models on structured tasks, where systematic gains at this scale are practically meaningful for autonomous driving applications.

\subsection{Proprietary Data and Reproducibility Scope}
\label{sec:reproducibility_scope}

Because some tasks are built from real-world autonomous-driving data, it is important to distinguish between findings that currently depend on that setting and findings that are readily reproducible on public resources. The least portable result is the \emph{capability-dependent reversal} in Table~\ref{tab:CoT_comparison}. Reproducing that phenomenon faithfully requires a controlled sequence of matched checkpoints within the same architecture and domain---from pre-alignment, to general driving capability, to object-specific adaptation, to task-level post-training---so that changes in CoT effectiveness can be attributed primarily to changing model capability rather than to changes in model family, task formulation, or data distribution. Existing public embodied benchmarks typically do not provide this combination of staged checkpoints, task-specific action supervision, and reasoning annotations in the same domain, so we do not claim that the exact quantitative pattern can already be reproduced end-to-end on open data.

By contrast, the methodological contribution of this paper is much easier to reproduce. Our structural probe is simply the comparison between think-then-act and decide-then-explain under an autoregressive model, which changes the causal position of CoT tokens without changing model weights or the underlying task. Any public VLA or VLM that supports CoT-style outputs can implement this probe by reordering the output format; with supervised reasoning data, the same factorization can also be trained directly. In this sense, the decide-then-explain paradigm is not tied to our dataset and should be straightforward to evaluate in other open embodied settings.

We therefore view the proprietary driving experiments as the setting that reveals the capability-dependence result most cleanly, rather than as the only evidence for our broader mechanistic interpretation. The appendix already includes several public-data supplements supporting portability of the main idea: OpenVLA experiments on LIBERO-90 (Appendix~\ref{sec:openvla_details}) reproduce the qualitative advantage of decide-then-explain in robotic manipulation, and zero-shot studies on public VLMs and benchmarks (Appendix~\ref{sec:vlm_details}) show that generation order can matter across model classes and training regimes. These open-source experiments do not replace the proprietary capability-scaling analysis, but they do show that both the structural-probe logic and the practical decide-then-explain recipe are readily reproducible beyond our driving dataset.

\section{Model Training Details}
\label{sec:training_detail}
\subsection{Training Details}

This section provides implementation-level details of model training, including hardware configuration, architecture design, optimization settings, and multi-stage training procedures.

\paragraph{Hardware and Compute.}
All models are trained using 256 NVIDIA H20 GPUs with 96GB memory per card. Training is conducted in a fully distributed setting with data parallelism to support large-scale vision-language-action learning.

\paragraph{Model Architecture.}
The vision-language-action model adopts a modular architecture consisting of a SigLIP-based~\citep{zhai2023sigmoidlosslanguageimage} visual encoder, a Qwen2.5-7B~\citep{qwen2025qwen25technicalreport, bai2025qwen25vltechnicalreport} decoder-only language model, and a lightweight MLP projector for visual-to-language feature alignment. The projector maps visual embeddings into the language model's token space. Unless otherwise specified, all modules—including the visual encoder, language model, and projector—are jointly optimized during training without freezing.

\paragraph{Optimization and Hyperparameters.}
Training is performed using the AdamW~\citep{loshchilov2019decoupledweightdecayregularization} optimizer with an initial learning rate of $5 \times 10^{-5}$ and a linear decay schedule. The batch size is 4 samples per GPU. A warmup ratio of 0.03 is applied at the beginning of training. All tasks are optimized using standard cross-entropy loss.

\subsection{Training Stages}

The VLA model is trained using a staged curriculum that progressively builds cross-modal alignment, general autonomous driving capabilities, object-specific understanding, and task-specific decision-making competence.

First, a projector pre-alignment stage is performed using the LLaVA-Pretrain dataset~\cite{liu2023visualinstructiontuning}. During this stage, the visual encoder and language model are frozen, and only the MLP projector is trained to establish basic vision-language alignment, resulting in a \textit{pre-aligned model}.

The model then enters a general capability learning phase consisting of two consecutive supervised fine-tuning (SFT) stages. In SFT Stage I, the model is trained on open-source autonomous driving datasets, including CC-OCR~\cite{qwen2025qwen25technicalreport}, Drive-LLM~\cite{DriveLLM}, SpatialSense~\cite{SpatialSense}, and MFE-ETP~\cite{MFE-ETP}, TDIUC~\cite{kafle2017analysis}, to acquire general visual understanding, instruction-following, and reasoning abilities in driving-related scenarios. This stage yields the \textit{AD-general model}.

In SFT Stage II, the model is further fine-tuned on professionally annotated datasets emphasizing critical traffic signs and semantic markers relevant to parking and low-speed navigation, producing the \textit{object-specific model}. Both SFT stages adopt full-parameter fine-tuning.

Finally, the object-specific model undergoes task-oriented post-training according to on vision-language-action datasets as described in Appendix \ref{sec:dataset_setup} for Navigation Decision and Target Path Generation, enabling decision-level adaptation for embodied tasks.

\section{Additional Evidence and Cross-Domain Evaluation Details}
\subsection{Evidence of Latent Reasoning}
\label{sec:latent_reasoning_proof}

\begin{figure}[htbp]
    \centering
    
    % 第一行：两个宽图
    \begin{subfigure}{0.48\textwidth}
        \includegraphics[width=\textwidth]{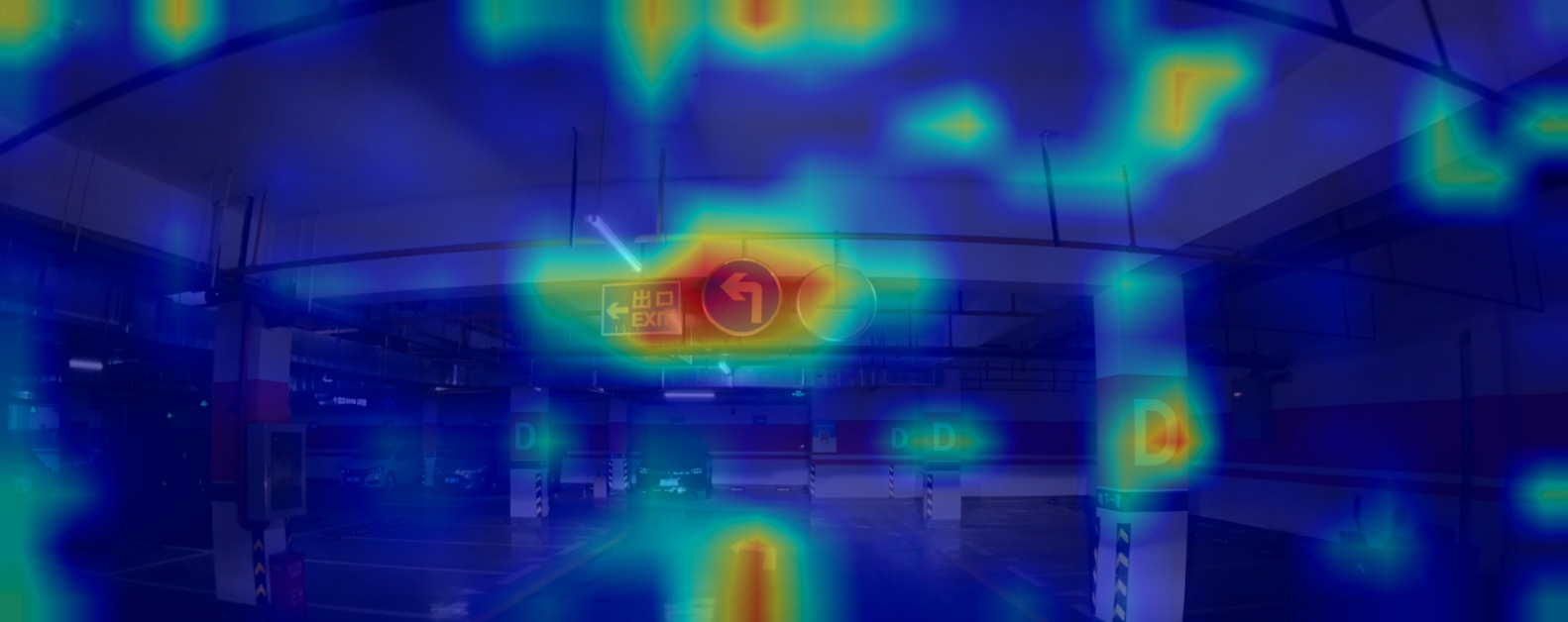}
        \caption{Navigation Decision with-CoT}
    \end{subfigure}
    \hfill
    \begin{subfigure}{0.48\textwidth}
        \includegraphics[width=\textwidth]{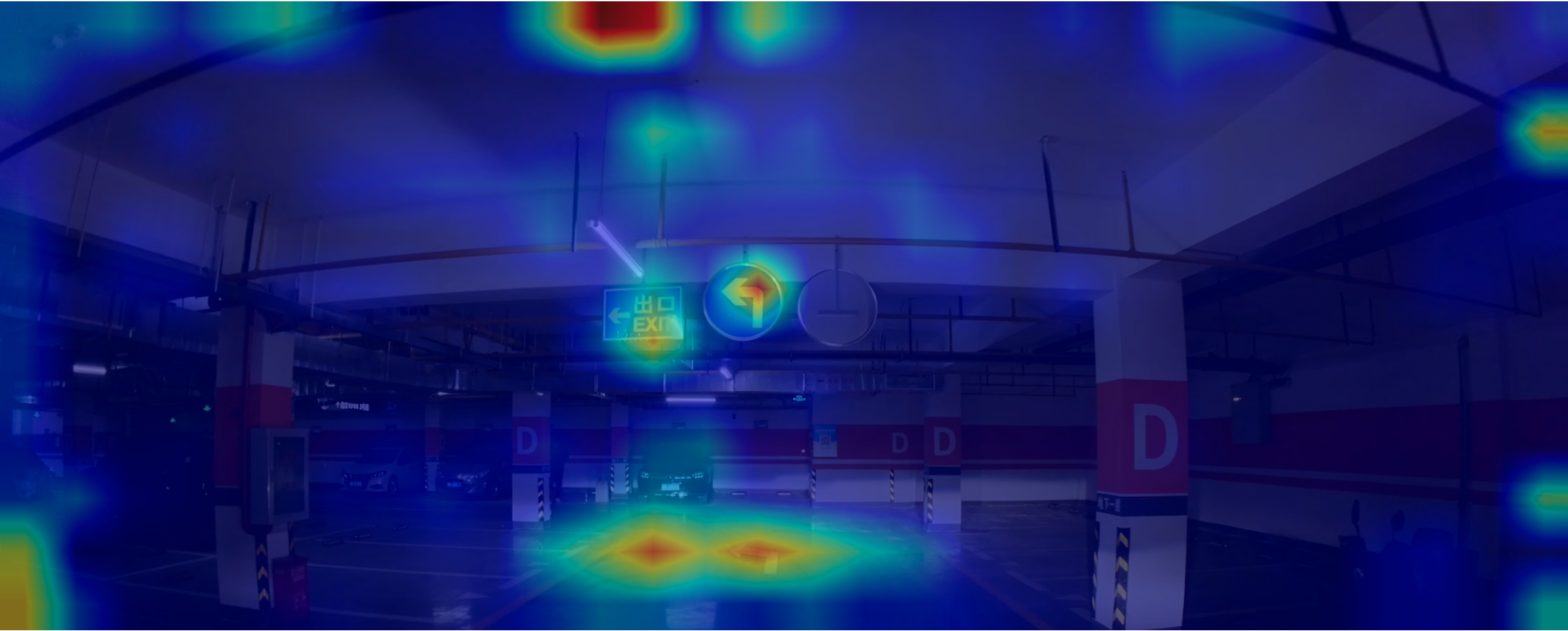}
        \caption{Navigation Decision without-CoT}
    \end{subfigure}
    
    % 第二行：两个窄图（注意：你第二个图路径重复了，应该是 without-CoT）
    \begin{subfigure}{0.48\textwidth}
        \includegraphics[width=\textwidth]{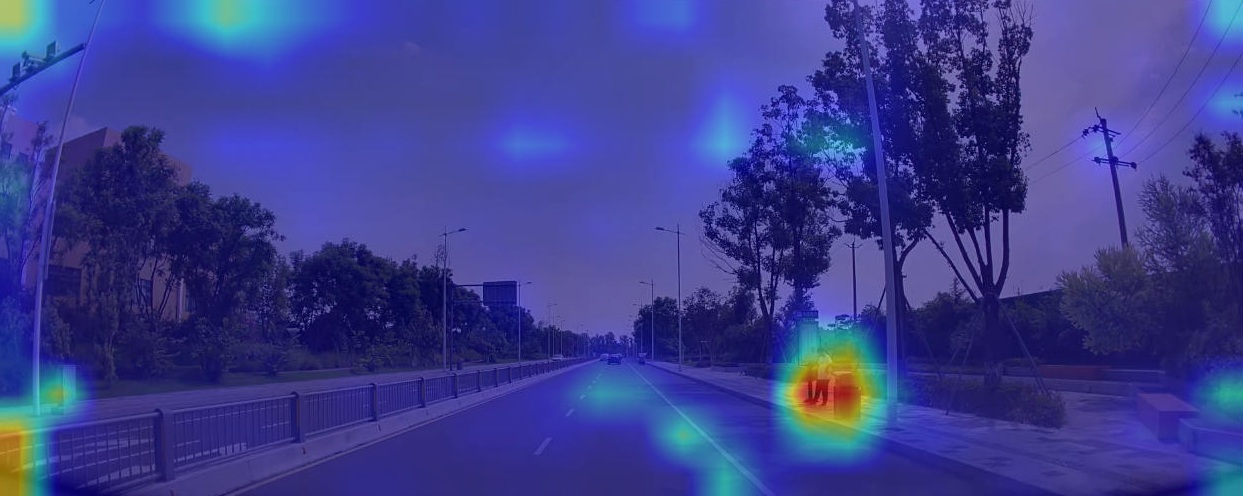}
        \caption{Target Path Generation with-CoT}
    \end{subfigure}
    \hfill
    \begin{subfigure}{0.48\textwidth}
        \includegraphics[width=\textwidth]{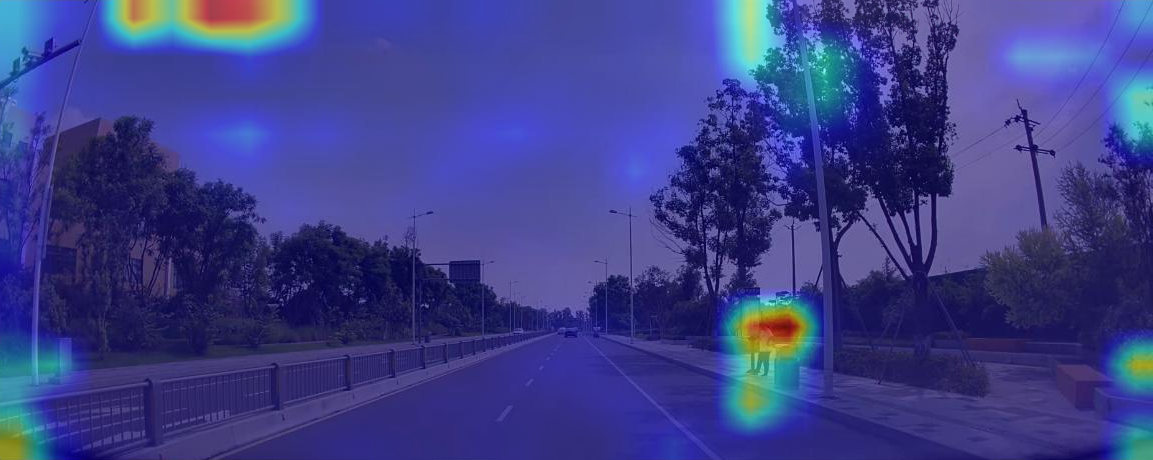}  % 修改了路径
        \caption{Target Path Generation without-CoT}
    \end{subfigure}

	\caption{Attention heat map visualizations under explicit (with-CoT) and latent (without-CoT) reasoning paradigms. Despite the absence of explicit reasoning traces, the without-CoT setting exhibits attention patterns concentrated on the same task-relevant visual regions, indicating preserved latent reasoning.}

	\label{fig:latent_reasoning_proof}
\end{figure}

This appendix provides empirical evidence that the model performs latent reasoning even in the absence of explicit CoT outputs.

In general, explicit reasoning manifests as textual references to task-critical semantic entities (e.g., key landmarks, traffic signs, or target objects), followed by a decision derived from these observations. Correspondingly, attention visualizations typically exhibit concentrated activations on these semantically relevant regions, which is commonly interpreted as evidence of reasoning behavior. In the \textit{without-CoT} setting, the model no longer produces intermediate textual reasoning or explicitly mentions task-relevant semantic cues. However, as shown in Fig.~\ref{fig:latent_reasoning_proof}, attention map visualizations reveal that the model still allocates high attention to the same task-critical visual entities as in the explicit reasoning setting. The spatial distribution and semantic focus of attention remain highly consistent across the two paradigms.

This observation indicates that, despite the absence of explicit reasoning traces in the output space, the model internally performs reasoning over visual information to support decision generation. The preservation of semantically aligned attention patterns provides strong evidence that reasoning is implicitly encoded within the shared representation space.

\subsection{OpenVLA Experiment Details}
\label{sec:openvla_details}

This section provides comprehensive implementation details for the OpenVLA experiments described in Section~\ref{sec:cross_domain}.

\subsubsection{Model Architecture}

We adopt the OpenVLA framework~\citep{kim24openvla} with a unified vision-language-action architecture. The model consists of three key components: (1) a Llama2-7B~\citep{touvron2023llama} decoder-only language model serving as the backbone, (2) dual vision encoders—DinoV2~\citep{oquab2023dinov2learningrobustvisual} and SigLIP~\citep{zhai2023sigmoidlosslanguageimage}—for visual feature extraction from third-person robot images, and (3) an action decoder that autoregressively predicts 7-dimensional continuous actions token-by-token. The action space includes delta translations ($\Delta x, \Delta y, \Delta z$), delta rotations ($\Delta \text{roll}, \Delta \text{pitch}, \Delta \text{yaw}$), and binary gripper state.

Following the original OpenVLA design, visual tokens from both encoders are projected into the language model's embedding space and concatenated with task instruction tokens. Unlike the baseline OpenVLA, which outputs only action tokens, we adapt the architecture to jointly generate CoT reasoning and action tokens in a unified autoregressive sequence.

\subsubsection{Training Configuration}

\paragraph{Hardware and Compute.}
Training is conducted on 32 NVIDIA H20 GPUs using LoRA-based parameter-efficient fine-tuning. The total training duration is 78 hours for 200,000 optimization steps.

\paragraph{Hyperparameters.}
We use a learning rate of $8 \times 10^{-4}$ with a batch size of 6 samples per GPU and no gradient accumulation (effective batch size: $6 \times 32 = 192$). Following OpenVLA's configuration, we employ the AdamW optimizer without learning rate warmup, training for a fixed number of steps rather than epochs.

\paragraph{LoRA Configuration.}
To enable efficient fine-tuning of the 7B-parameter model, we apply Low-Rank Adaptation (LoRA)~\citep{hu2021loralowrankadaptationlarge} with rank $r=32$.

\subsubsection{Dataset and Preprocessing}

We use the LIBERO-90~\citep{liu2023liberobenchmarkingknowledgetransfer} reasoning dataset used by ECoT-Lite~\citep{chen2025trainingstrategiesefficientembodied}, which provides 90 diverse robotic manipulation tasks. The dataset comprises 3,917 successful demonstration trajectories (approximately 50 trajectories per task) collected by rolling out actions from the original LIBERO release and filtering unsuccessful executions.

Each trajectory includes third-person RGB observations, natural language task instructions, ground-truth 7-DoF actions, and structured CoT annotations. The CoT annotations follow a hierarchical format specifying high-level plans, subtask decomposition, movement reasoning, object bounding boxes, and gripper position, enabling the model to learn structured reasoning processes.

\subsubsection{Evaluation Protocol}

We follow the standard LIBERO-90 evaluation protocol. Each of the 90 tasks is evaluated over 10 independent trials, and success rates are computed as the percentage of successful task completions. A trial is considered successful when the task goal is achieved within the maximum episode length of 400 steps, following the same success criteria as the original OpenVLA implementation.

The evaluation environment uses the MuJoCo~\citep{todorov2012mujoco} physics simulator with a third-person camera configuration. To ensure valid comparisons, we filter out anomalous episodes where the robot remains completely stationary (all actions are zero vectors) for the entire 400-step duration, as these represent failure cases due to ``no-op'' prediction artifacts rather than meaningful policy behavior.

\subsubsection{Example Task Outputs}

We provide representative examples illustrating the model's behavior under different reasoning paradigms.

\begin{quote}
\noindent\textbf{Task Instruction:} ``Pick up the book on the left and place it on top of the shelf.''

\smallskip
\noindent\textbf{Visual Input:} Third-person view of tabletop scene with books and a two-layer wooden shelf.

\smallskip
\noindent\textbf{Decide-then-Explain Output:}

\texttt{<answer>}[7-DoF action tokens]\texttt{</answer>}

\texttt{<think>}plan: move to the left | grasp the book | move the book away from the shelf | move to the shelf | place the book on top of the shelf | release the book | subtask: move the book away from the shelf | subtask\_reasoning: the robot needs to move the book away from the shelf to create space and a clear path to reach and place it on top of the shelf | movement: move up and right and forward | movement\_reasoning: the shelf is positioned in the upper right corner of the image, and the book is on the left side, so the robot needs to move diagonally upward and to the right to reach the shelf | bboxes: black book 1 [[124, 58], [177, 75]], yellow book 1 [[120, 97], [164, 123]], wooden two layer shelf 1 [[80, 164], [196, 223]] | gripper: [73, 44]\texttt{</think>}

\smallskip
\noindent\textbf{Think-then-Act Output:}

\texttt{<think>}[identical reasoning content as above]\texttt{</think>}

\texttt{<answer>}[7-DoF action tokens]\texttt{</answer>}
\end{quote}

\subsection{VLM Zero-Shot Generalization Details}
\label{sec:vlm_details}

This section provides full experimental details for the zero-shot VLM generalization study described in Section~4.3.4.

\subsubsection{Benchmarks}

\paragraph{MMVU.} A multidisciplinary video question-answering benchmark requiring integration of visual perception and domain knowledge.

\paragraph{VSI-Bench.} A video spatial intelligence benchmark testing 3D spatial reasoning from egocentric video observations, including object counting, spatial relations, and distance estimation.

\paragraph{VideoMME.} A comprehensive video multi-modal evaluation benchmark spanning diverse video types and durations, testing general-purpose video understanding capabilities.

\subsubsection{Models}

\paragraph{Qwen2.5-VL-7B.} A 7B-parameter vision-language model from the Qwen series~\citep{bai2025qwen25vltechnicalreport}, trained on large-scale image and video data with strong instruction-following capabilities. We use the instruction-tuned variant without any further fine-tuning.

\paragraph{Video-R1.} A video understanding model trained with reinforcement learning~\citep{videor1}, Video-R1 is enhanced for multi-step reasoning over video inputs. Its strengthened reasoning capabilities allow it to perform complex inference across temporal sequences, making it particularly suitable for studying generation-order effects.

\subsubsection{Prompting Strategy}

For both models, we apply prompts that differ only in the prescribed output ordering:

\textbf{Think-then-act:} The model is instructed to first produce a reasoning trace enclosed in \texttt{<think>...</think>}, followed by the final answer in \texttt{<answer>...</answer>}.

\textbf{Decide-then-explain:} The model is instructed to first produce the final answer in \texttt{<answer>...</answer>}, followed by the explanatory reasoning in \texttt{<think>...</think>}.

All other aspects of the prompt---task description, answer format, and evaluation criteria---are held constant. No model parameters are updated; the paradigm shift is purely at the inference-time token ordering level.

\section{Causal Intervention Details}
\label{sec:causal_interference_details}
\subsection{Causal Interference Experiment Details}

This section provides full experimental details for the CoT causal interference study described in Section~4.3.

\subsubsection{Setup}

We use the Object-Specific model trained under the think-then-act paradigm. At inference time, we intercept the model's \texttt{<think>...</think>} block and replace its content with a fixed, pre-written CoT before the model generates its \texttt{<answer>}. All other inputs---visual observations, task instructions, and model parameters---remain identical to the standard evaluation. This constitutes a controlled intervention in the interventional (do-calculus) sense: we directly set the value of the CoT variable while holding everything else constant. Any change in decision output can therefore be attributed causally to the CoT content, not to confounding factors.

\subsubsection{Corrupted CoT Content}

\paragraph{Navigation Decision.}
The fixed CoT injected for all Navigation Decision samples is sampled from the model's own generated CoT outputs, ensuring statistical consistency with the training distribution. The injected trace reads:

\begin{quote}
\textit{I can see a right-turn exit sign in the scene. The road ahead curves to the right, and the exit lane is clearly marked on the right side. Based on the visual cues, the correct action is to follow the right-turn exit. Navigation decision: turn right.}
\end{quote}

This CoT unconditionally encodes a right-turn conclusion regardless of the actual visual content, which may show straight paths, left turns, or other directions.

\paragraph{Target Path Generation.}
The fixed CoT injected for all Target Path Generation samples is sampled from the model's own generated CoT outputs, ensuring statistical consistency with the training distribution. The injected trace reads:

\begin{quote}
\textit{The road ahead is lined with trees on both sides. A silver SUV is parked on the right side of the road approximately 15 meters ahead. The road surface is dry and clear. To reach the target, the vehicle should steer right-forward toward the silver SUV, adjusting the trajectory to pass safely on the left side.}
\end{quote}

This CoT unconditionally describes a specific static scene and prescribes a right-forward steering direction, regardless of the actual environment or designated target location.

\subsubsection{Why This Is a Valid Causal Test}

The standard correlation-based analysis (e.g., comparing models trained with versus without CoT) cannot distinguish whether CoT \emph{causes} decisions or merely co-occurs with them due to shared training signal. Our intervention breaks this ambiguity. By surgically replacing only the CoT content at inference time---without modifying visual input, instructions, or weights---we isolate the causal effect of the CoT token stream on the decision output. The dramatic collapse in Navigation F1 (from 72.14\% to 11.84\%) and the consistent degradation in Target Path Generation metrics (ADE: 1.83\,m $\to$ 2.00\,m; FDE: 3.17\,m $\to$ 3.48\,m) confirm that the model does not override the injected reasoning with independent visual evidence. Under think-then-act inference, the CoT is a causal driver of the decision tokens that follow it.

\subsubsection{Plausible-but-Wrong CoT Intervention }
\label{sec:plausible_wrong_cot}

\paragraph{Motivation.}
The fixed-trace intervention above establishes that CoT causally controls decisions, but leaves open a lexical-matching alternative: the model may simply copy directional tokens (e.g., ``turn right'') from the injected trace without genuinely following its reasoning conclusion. To rule this out, we design a second intervention that preserves the semantic plausibility of the CoT while inverting only its directional conclusion.

\paragraph{Procedure.}
We run inference twice. In the first pass, the model generates a complete CoT from the visual input; this trace is scene-grounded and semantically consistent with the observation. In the second pass, we replace only the directional decision tokens within the trace---e.g., ``turn left'' $\to$ ``turn right'', ``left side'' $\to$ ``right side''---while leaving all scene descriptions, spatial analysis, and logical connectives intact. The modified trace is then used as the inference-time CoT for the second forward pass.

\paragraph{Coverage.}
Navigation Decision CoT traces contain directional tokens in virtually every sample (left/right/straight appear in the decision conclusion), so intervention coverage is near-complete. Target Path Generation CoT traces contain directional tokens less consistently, as waypoint reasoning may describe spatial relationships without explicit left/right vocabulary; intervention coverage is therefore partial for this task.

\paragraph{Examples.}

\textit{Navigation Decision --- original trace (excerpt):}
\begin{quote}
\ldots\ Navigation decision: based on the left-turn elevator sign and the intersection analysis, the navigation decision is to \textbf{turn left}. Turning left will lead directly into the corridor toward the elevator hall.
\end{quote}

\textit{After directional inversion:}
\begin{quote}
\ldots\ Navigation decision: based on the left-turn elevator sign and the intersection analysis, the navigation decision is to \textbf{turn right}. Turning right will lead directly into the corridor toward the elevator hall.
\end{quote}

\textit{Target Path Generation --- original trace (excerpt):}
\begin{quote}
\ldots\ Navigation decision: to park near the target vehicle, the vehicle should drive toward the \textbf{right-front}, approach and stop beside the black luxury car on the \textbf{right} side.
\end{quote}

\textit{After directional inversion:}
\begin{quote}
\ldots\ Navigation decision: to park near the target vehicle, the vehicle should drive toward the \textbf{left-front}, approach and stop beside the black luxury car on the \textbf{left} side.
\end{quote}

\paragraph{Results and Interpretation.}
Navigation F1 collapses to 16.8\% (vs.\ 11.84\% under the fixed-trace intervention), and Target Path Generation degrades to ADE 1.88\,m / FDE 3.27\,m (vs.\ 2.00\,m / 3.48\,m). The slightly smaller degradation relative to the fixed-trace intervention is consistent with two factors: (i) the scene descriptions in the plausible-but-wrong CoT remain consistent with the visual input, providing a marginal corrective signal; and (ii) intervention coverage on Target Path Generation is partial due to lower directional token density. Crucially, the near-identical order-of-magnitude collapse across both interventions confirms that the model is not merely performing lexical matching on directional tokens---it is genuinely following the reasoning conclusion encoded in the CoT.

\section{Latency Analysis of Generation Paradigms}
\label{sec:latency_analysis}
\subsection{Latency Analysis of Generation Paradigms}

This section analyzes the inference latency implications of the decide-then-explain and think-then-act paradigms, and demonstrates that decide-then-explain affords a structural latency advantage in latency-sensitive deployment settings.

\paragraph{Theoretical Equivalence Under Fixed Output Length.}
Autoregressive language model inference generates one token per forward pass; thus, end-to-end generation latency scales linearly with the total number of output tokens. Let $T_d$ denote the number of decision tokens (i.e., the action or answer) and $T_r$ denote the number of reasoning tokens (i.e., the CoT explanation). When both paradigms produce identical token sequences differing only in generation order, the total token count is $T_d + T_r$ in both cases, and the end-to-end latencies are theoretically equivalent.

\paragraph{Practical Asymmetry via Early Truncation.}
In practice, the necessity of explicit reasoning varies across deployment scenarios. The structural difference between the two paradigms creates a fundamental asymmetry in how early truncation can be applied:

\begin{itemize}
    \item \textbf{Decide-then-explain:} The generation order is $\langle\texttt{answer}\rangle \cdots \langle\texttt{/answer}\rangle \langle\texttt{think}\rangle \cdots \langle\texttt{/think}\rangle$. Decision tokens are produced first. Inference can be halted immediately upon closing the \texttt{</answer>} tag, discarding the subsequent reasoning entirely. The latency to obtain the action is $\mathcal{O}(T_d)$, independent of $T_r$.

    \item \textbf{Think-then-act:} The generation order is $\langle\texttt{think}\rangle \cdots \langle\texttt{/think}\rangle \langle\texttt{answer}\rangle \cdots \langle\texttt{/answer}\rangle$. The model must complete the full reasoning trace before the decision token becomes available. The minimum latency to obtain the action is $\mathcal{O}(T_r + T_d)$, regardless of whether the downstream application requires the reasoning.
\end{itemize}

Since CoT annotations are substantially longer than action sequences in embodied tasks ($T_r \gg T_d$), the effective speedup of decide-then-explain under early truncation is approximately:
\begin{equation}
    \text{Speedup} = \frac{T_r + T_d}{T_d} = 1 + \frac{T_r}{T_d}.
\label{eq:speedup}
\end{equation}

In our experimental setting, CoT sequences average \textbf{130 tokens} (max 180) for autonomous driving tasks, with decision token sequences $T_d = 3$ tokens for navigation commands, yielding a practical speedup of $\frac{130 + 3}{3} \approx 44\times$ under early truncation. For robotic manipulation (OpenVLA), CoT sequences average \textbf{172 tokens} (max 258) with $T_d = 7$ action tokens, yielding a speedup of $\frac{172 + 7}{7} \approx 26\times$. These results confirm that the latency advantage of decide-then-explain is substantial across embodied domains, particularly in settings where real-time responsiveness is essential.

\paragraph{Deployment Flexibility.}
This structural property enables the decide-then-explain paradigm to adaptively trade off latency against interpretability at inference time, without any modification to the underlying model weights:

\begin{itemize}
    \item \textbf{Latency-critical scenarios} (e.g., real-time autonomous driving or reactive robot control): Generation is truncated after the action token sequence, achieving latency comparable to the without-CoT baseline while retaining the full performance benefits conferred by decide-then-explain training.

    \item \textbf{Interpretability-required scenarios} (e.g., post-hoc safety audits, operator oversight, or debugging): The model generates the complete output, providing both the action and its accompanying reasoning explanation.
\end{itemize}

In contrast, think-then-act does not support this form of adaptive inference: suppressing the reasoning phase entirely would require prompting the model to omit its \texttt{<think>} block, effectively switching to a without-CoT mode and foregoing the reasoning-conditioned generation pathway. The decide-then-explain paradigm thus provides a unified inference interface that subsumes both the without-CoT (fast) and with-CoT (interpretable) operating modes under a single trained model.

\section{Evidence for Indirect Training Benefits of CoT}
\label{sec:indirect_learning_ablation}
\subsection{Indirect Parameter Learning: Consolidated Comparison}

This subsection consolidates key performance data to support the claim that CoT annotations benefit decision learning indirectly through gradient-level parameter updates during training, rather than by directly influencing decision computations at inference time.

Table~\ref{tab:indirect_learning} compares the w/o CoT baseline against representative CoT-trained configurations for the Navigation Decision task (Object-Specific backbone). In the decide-then-explain setting, the model generates its action tokens before producing any reasoning trace; consequently, the explicit reasoning coefficient $c_r \approx 0$, confirming that CoT tokens exert no direct computational effect on the decision output at inference time. Despite this, decide-then-explain (74.50\%) consistently outperforms the w/o CoT baseline (73.49\%) by +1.01 percentage points in Navigation F1. Under think-then-act training, explicit reasoning does influence decisions at inference time ($c_r = 1.81$), yet performance does not surpass the w/o CoT baseline, underscoring that the performance gain in decide-then-explain cannot be attributed to direct CoT-driven inference.

The only remaining mechanism through which CoT annotations could produce this improvement is indirect: during training, CoT supervision shapes the model's parameters via gradient updates, improving the quality of latent representations and visual grounding even in the absence of any reasoning signal at test time.

\begin{table}[h]
\centering
\caption{Consolidated comparison of training strategies for Navigation Decision (Object-Specific backbone). The w/o CoT baseline contains no CoT structure; $c_v$/$c_r$ are not applicable (---).}
\label{tab:indirect_learning}
\small
\setlength{\tabcolsep}{6pt}
\renewcommand{\arraystretch}{1.1}
\begin{tabular}{llccccc}
\toprule
\textbf{Training Strategy} & \textbf{Inference} & \textbf{Nav F1 (\%) $\uparrow$} & \textbf{ADE (m) $\downarrow$} & \textbf{FDE (m) $\downarrow$} & $\mathbf{\bar{c_v}}$ & $\mathbf{\bar{c_r}}$ \\
\midrule
w/o CoT (baseline)   & ---                  & 73.49          & 1.82          & 3.16          & ---  & ---  \\
Think-then-act       & Think-then-act       & 72.14          & 1.83          & 3.17          & 0.25 & 1.81 \\
Decide-then-explain  & Decide-then-explain  & \textbf{74.50} & \textbf{1.80} & \textbf{3.13} & 1.27 & 0.00 \\
\bottomrule
\end{tabular}
\end{table}

Note that $c_v$ and $c_r$ values are not reported for the w/o CoT condition because that model produces no CoT structure, making the coefficient decomposition undefined. The core comparison therefore rests on the F1/ADE/FDE metrics.

\subsection{Task-Wise Differences in Contribution Metrics}

The contribution metrics also reveal systematic differences between the two tasks that reflect their intrinsic reliance on language-mediated versus perception-grounded computation. Navigation Decision shows high $c_r$ and low $c_v$ under think-then-act, indicating stronger dependence on explicit reasoning and correspondingly greater susceptibility to CoT interference. Target Path Generation maintains comparatively high $c_v$ across paradigms, reflecting its greater reliance on fine-grained spatial perception and lower sensitivity to reasoning-path placement. These task-wise patterns are consistent with the nature of each task: navigation direction prediction depends heavily on scene-level semantic interpretation, whereas waypoint generation requires precise spatial grounding that is more directly encoded in visual features.

\section{Improved CoT Design and Format Analysis}
\label{sec:improved_cot_details}
\subsection{Improved CoT: Format Analysis and Design Principles}

\begin{figure}[h]
\centering
\includegraphics[width=1 \textwidth]{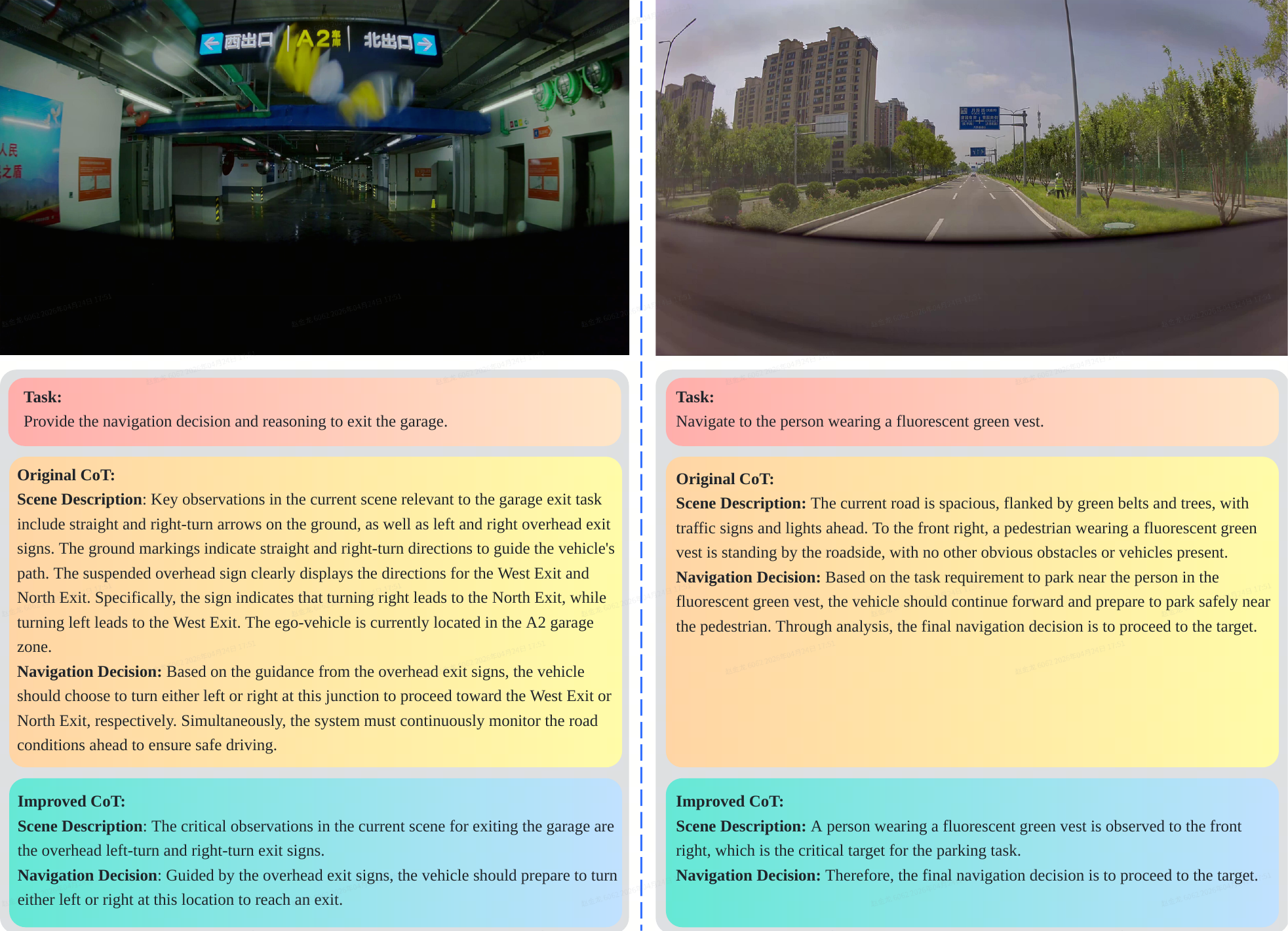}
\caption{Comparison of Original CoT (verbose, scene-centric) and Improved CoT (concise, causal). Original CoT enumerates task-irrelevant features (zone labels, background objects, generic safety prompts), introducing textual noise that can interfere with decision gradients. Improved CoT focuses on task-relevant observations with direct causal links to the decision.}
\label{fig:improved_cot}
\end{figure}

This appendix provides both a procedural and qualitative comparison between Original CoT and Improved CoT formats, explaining how the improved data are constructed and why the resulting reasoning traces have higher quality. The Improved CoT is generated from expert-guided structural annotations rather than by unconstrained free-form description alone. Fig.~\ref{fig:improved_cot} illustrates the key format differences, while the following subsections describe the generation pipeline and corpus-level statistics.

\subsubsection{Scene Extraction and Generation Pipeline}

Both CoT generation pipelines share a common upstream stage: a general-purpose scene recognition model is first applied to each input image to extract task-relevant key scenes. These extracted cues serve as structured perceptual inputs for downstream reasoning generation, ensuring that Original CoT and Improved CoT are grounded on the same visual primitives.

\paragraph{Original CoT Generation.}
The Original CoT is produced by providing the task instruction, raw input image, and extracted task-relevant key scenes to a large multimodal language model, which then directly performs open-ended scene analysis and decision reasoning. No additional structural constraints or expert guidance are imposed, so the model freely generates a relatively comprehensive scene description followed by its reasoning trace.

\paragraph{Improved CoT Generation.}
The Improved CoT is generated through a two-stage guided pipeline:
\begin{enumerate}
    \item \textbf{Expert annotation construction.} The extracted task-relevant key scenes are first organized into a scene-level spatial map, from which the decision is derived through rule-based spatial reasoning. The resulting concise logic links specific visual cues to their decision implications, yielding expert annotations that explicitly encode the causal chain from perception to action.
    \item \textbf{Guided natural-language generation.} The task instruction, raw input image, extracted key scenes, and expert annotations are then jointly provided to a large multimodal language model, which refines these annotations into a natural-language CoT while preserving a tight logical structure from scene perception through scene association to decision reasoning.
\end{enumerate}

This pipeline differs from Original CoT in a crucial way: instead of asking the language model to freely narrate the scene and reason from scratch, we first supply a compact expert scaffold that constrains the generated CoT to remain causally aligned with the task-relevant perceptual evidence.

\subsubsection{Quantitative Comparison}

To quantitatively characterize the differences between Original CoT and Improved CoT, we compute corpus-level statistics over the full training set. Specifically, we measure the average character count of the CoT content, then use a language model-based parser to identify task-relevant scenes, task-irrelevant scenes, and generic safety-related prompts in each CoT instance. We report the task-relevant scene retention rate, the average number of task-irrelevant scenes per sample, and the safety-prompt occurrence rate. Results are summarized in Table~\ref{tab:cot_quantitative}.

\begin{table}[h]
\centering
\caption{Quantitative comparison between Original CoT and Improved CoT. Task-relevant scene retention measures completeness; task-irrelevant scene count and safety-prompt rate measure textual noise.}
\label{tab:cot_quantitative}
\small
\setlength{\tabcolsep}{10pt}
\renewcommand{\arraystretch}{1.1}
\begin{tabular}{lcc}
\toprule
\textbf{Metric} & \textbf{Original CoT} & \textbf{Improved CoT} \\
\midrule
Average character count            & $\sim$130 & $\sim$75 \\
Task-relevant scene retention rate & 100\%     & 100\%    \\
Task-irrelevant scenes (avg.)      & 4.2        & 0.3       \\
Safety prompt occurrence rate      & 85\%      & 0\%      \\
\bottomrule
\end{tabular}
\end{table}

As shown in Table~\ref{tab:cot_quantitative}, the Improved CoT preserves full coverage of task-relevant scenes while substantially reducing noise: task-irrelevant scene mentions decrease from 4.2 to 0.3 per sample, generic safety prompts are eliminated, and the overall CoT length is reduced by approximately 42\%. These statistics complement the qualitative examples by showing that the improved format is not merely shorter, but also more selective and causally focused.

\paragraph{Original CoT: Verbose, Panoramic Description.}
The Original CoT employs a descriptive, scene-centric approach that catalogs many environmental features. For garage navigation tasks, it enumerates elements such as:
\begin{itemize}
    \item Traffic arrows on the ground (e.g., ``straight and right-turn arrows'')
    \item Zone labels (e.g., ``A2 garage zone,'' ``Zone C1'')
    \item Pillar markings and background objects
    \item Generic scene descriptions (``several parked vehicles ahead,'' ``no pedestrians present'')
    \item Redundant safety prompts (``continuously monitor road conditions to ensure safe driving'')
\end{itemize}

\paragraph{Improved CoT: Causal, Task-Oriented Structure.}
In contrast, the Improved CoT focuses on task-relevant information and preserves only the direct causal links between visual observations and decisions. The design principles are:

\begin{enumerate}
    \item \textbf{Task-relevant only:} Isolate observations that directly influence the decision (e.g., ``overhead right-turn exit sign'' instead of comprehensive scene cataloging).

    \item \textbf{Reduce generic descriptions:} Remove zone labels, background object inventories, and environmental context unless they inform the immediate action.

    \item \textbf{Stronger causal mapping:} Structure reasoning as `observation $\to$ interpretation $\to$ decision', explicitly linking each key observation to its role in determining the action.

    \item \textbf{Remove generic safety prompts:} Omit boilerplate advice that provides no decision-specific constraint.
\end{enumerate}

\end{document}